# Early-age formation of MgO-$SiO_2$ binders and the effect of hydromagnesite and dehydrated hydromagnesite: An in situ synchrotron X-ray diffraction and pair distribution function study

Jordan Mravca-Bailey[a,b], Claire E. White[a,b*]

[a] Department of Civil and Environmental Engineering, Princeton University, USA

[b] Andlinger Center for Energy and the Environment, Princeton University, USA

[*]Corresponding author.

**Abstract**

MgO-$SiO_2$ binders are promising low-$CO_2$ alternatives to Portland cement that can permanently store $CO_2$ by incorporating magnesium carbonate phases. Adding a magnesium carbonate such as nesquehonite or hydromagnesite is known to change the initial hydration product from crystalline brucite ($Mg(OH)_2$) to disordered hydrous carbonate-containing brucite (HCB), yet how these binders evolve during the initial hours of reaction remains largely unknown. Here, in situ synchrotron X-ray diffraction (XRD) and pair distribution function (PDF) analysis follows an MgO-$SiO_2$ paste with and without hydromagnesite (HY, $Mg_5(CO_3)_4(OH)_2{\cdot}4H_2O$) or dehydrated hydromagnesite (DHY, HY heated at 325 °C to remove water) over the first 24 h, with ex situ measurements to 42 days. Real space phase quantification reveals that magnesium-silicate-hydrate (M-S-H) gel is detected in the first scan (~45 min) of every paste, before formation of any brucite or HCB. Hence, early-age M-S-H can precipitate directly from dissolving MgO and $SiO_2$, in contrast with the commonly described reaction sequence in which M-S-H forms only after brucite or HCB has precipitated. The PDF results also show that in the neat MgO-$SiO_2$ paste, half of the M-S-H present at 42 days has formed by 24 h, while no M-S-H is detected by XRD at any age due to the intrinsic disorder present in this phase.

Both carbonate sources (HY and DHY) are found to increase the extent of MgO dissolution, where no unreacted MgO is detected at 24 h in contrast with >7 days for the neat paste. However, DHY is seen to slow the initial stages of MgO dissolution (prior to ~6 h) compared with the neat paste, a period during which significant DHY rehydration is seen to occur without any HCB formation. Hence, this investigation shows that the onset of HCB precipitation in MgO-$SiO_2$-(D)HY paste is influenced by the hydration state of the DHY phase, where precipitation is delayed while DHY is rehydrating.

## 1. Introduction

Portland cement production accounts for 7-8% of global anthropogenic $CO_2$ emissions, with 60-70% of these emissions coming from the unavoidable release of $CO_2$ during calcination of limestone ($CaCO_3$).[1,2] Global demand for cement is increasing, projected to reach up to 6 Gt/yr by 2050.[3,4] Given the sustainability concerns regarding Portland cement usage, cement binders based on reactive MgO are an appealing alternative since MgO can be sourced from Mg-rich brines or from the dissolution of Mg-silicate rocks,[5,6] neither of which involves decomposition of a carbonate phase. The processing of Mg-silicate rocks also yields amorphous $SiO_2$, which can be used in conjunction with reactive MgO to form MgO-$SiO_2$ binders.[7] Binders based on reactive MgO, with or without the inclusion of amorphous $SiO_2$, can additionally incorporate magnesium carbonate, which offers a route for permanent storage of $CO_2$ within the material.[8–10]

The two hydration products of a binder made from reactive MgO and amorphous $SiO_2$ are brucite ($Mg(OH)_2$) and magnesium silicate hydrate (M-S-H).[11–13] The accepted reaction sequence is that MgO hydrates rapidly to brucite, which then reacts with dissolved silicates (from dissolution of amorphous $SiO_2$) to form M-S-H, the main strength-bearing phase in this system.[5,7,14–16] The second step is slow because brucite dissolution is kinetically hindered by the dissolved silicates.[17,18] At room temperature the transformation has been reported as incomplete, based on the presence of residual brucite, for Mg/Si molar ratio greater than 1.2 even after two years of hydration.[11]

The understanding of this reaction sequence has been assembled almost entirely from ex situ measurements, while the initial 24 h of reaction remains poorly resolved. From ex situ measurements of pastes made from MgO and silica fume (Mg/Si molar ratio = 1, w/s wt. ratio = 2, 20 °C) at 1 day, Tonelli et al. found using X-ray diffraction (XRD), $^{29}Si$ magic angle spinning nuclear magnetic resonance ($^{29}Si$ MAS NMR), and thermogravimetric analysis (TGA), that almost all of the MgO had hydrated to brucite (determined using XRD), ~40% of the silica had been consumed (determined using $^{29}Si$ MAS NMR), and M-S-H was present (determined using $^{29}Si$ MAS NMR, TGA, XRD).[13] In pastes of MgO and silica fume with a lower water content and the addition of 1 wt% sodium hexametaphosphate (Mg/Si = 1, w/s = 0.5 and 0.8), Zhang et al. found by TGA measurements only a trace amount of M-S-H at 2 days, with the majority of M-S-H forming after 2 days.[14] Isothermal calorimetry by Li et al. of pastes containing MgO and silica fume (Mg/Si = 2.2, w/s = 1, 25 °C) showed the first 24 h was dominated by the exotherm from brucite formation, followed by a shoulder attributed to M-S-H formation from 60 to 240 h.[19] This same investigation attributed a small amount of the first 24 h of heat release to M-S-H formation, consistent with the 50 to 150 °C TGA mass loss at 1 and 3 days.[19] Measurements within the first day have been made on pastes of MgO and microsilica (Mg/Si = 2.25, w/s = 2, 50 °C), with samples taken at 6, 12, and 24 h, soaked in ethanol for 24 h to stop hydration, then dried at 50 °C for 24 h before XRD.[20] In these samples, Zhang et al. found that brucite formed within the first 6 h and remained the main hydrate, whereas adding 10 wt% sodium metasilicate suppressed brucite formation and produced M-S-H by 24 h, which they attributed to the elevated dissolved magnesium and silicate concentrations.[20]

No investigation of $MgO-SiO_2$ pastes has followed the individual phases in situ during the first few hours of reaction, so it is not known whether M-S-H forms only through the reaction of brucite with silica or also directly from the dissolving MgO. A recent RILEM review of $Mg^{2+}$ dissolution and precipitation in MgO-based cements outlined the need for time-resolved characterization of M-S-H precipitation, and of the effect of additives such as carbonates, under cement-relevant conditions.[21] Synchrotron XRD and pair distribution function (PDF) are well suited to closing this research gap since they allow in situ data to be collected on the timescale of minutes,[22] therefore making it possible to probe the behavior of these pastes during the early hours of the reaction.[23,24] These two techniques complement each other since XRD (where data are analyzed in reciprocal space) provides insight on the crystalline phases while PDF analysis (data in real space) is sensitive to all phases, including the local structure of poorly ordered phases such as M-S-H.

Hydrated magnesium carbonates such as nesquehonite ($MgCO_3 \cdot 3H_2O$) and hydromagnesite (HY, $Mg_5(CO_3)_4(OH)_2 \cdot 4H_2O$) can be added to store $CO_2$ within the MgO-based binder, and they change both the products and the kinetics of the reaction. When a hydrated magnesium carbonate was included in a paste of reactive MgO, with or without amorphous $SiO_2$, hydrous carbonate-containing brucite (HCB) formed in place of crystalline brucite.[25–28] HCB is a poorly crystalline layered phase with a brucite-like framework that takes up (bi-)carbonate anions and loosely bound water molecules in its interlayer space, thereby altering the spacing between the hydroxide sheets and disrupting their stacking.[29] The carbonate and water content of HCB vary with the carbonate source and curing temperature.[29] In systems that contained only reactive MgO and magnesium carbonate, HCB has been found to be the principal binding phase.[30–32] Isothermal calorimetry of MgO-HY pastes showed that HY accelerates the early hydration of MgO, moving the second heat flow maximum from ~8 h to ~6 h.[25] When amorphous $SiO_2$ is also present, HCB acts as a metastable precursor phase for M-S-H and is seen to react over the course of weeks to months.[27,29] These conclusions have relied on ex situ measurements at one day or longer and on calorimetry, which by itself cannot reveal how individual phases react and contribute to the measured heat flow. Hence, the timing of HCB formation relative to MgO dissolution and M-S-H formation within the first day remains unknown, as does whether the early M-S-H forms through HCB or directly from MgO dissolution.

Even less is known about the effect of dehydrating the magnesium carbonate phase before incorporating it into the binder. Heating HY at 325 °C removes its crystalline water while leaving the carbonate groups intact, creating a disordered dehydrated hydromagnesite (DHY) phase.[33] The motivation for using DHY comes from work on thermally activated nesquehonite. Morrison et al. found that nesquehonite heated between 100 and 250 °C shows self-cementing behavior upon being mixed with water, generating cubes with compressive strength of ~2-4 MPa.[34] We hypothesize that DHY may behave in a similar way within the binder, adding to the cohesion and strength of the paste. Botha and Strydom reported that hydromagnesite heated at the same temperature returns to crystalline HY after 11 weeks in saturated water vapor, although the reflections in their XRD patterns remain broader than those of the starting material.[35] Substituting DHY for HY therefore raises two questions that this work addresses: how DHY behaves in a paste where MgO competes for available water, and whether the disorder introduced by dehydration affects the resulting reaction products.

Here we combine in situ synchrotron XRD and PDF analysis to investigate how HY and DHY affect the phase evolution behavior of an $MgO-SiO_2$ binder over the initial 24 h of reaction, with ex situ measurements extending to 42 days. The kinetics of MgO dissolution, brucite/HCB formation, and rehydration of DHY (for the DHY-containing paste) are tracked using XRD, while the impact of HY and DHY on the crystallinity of the brucite-type phase is assessed by quantification of XRD peak broadening. Real space phase quantification of the PDF data is used to determine the kinetics of M-S-H formation along with providing complementary kinetics insight on MgO dissolution and brucite/HCB formation. Combining these two techniques allows the sequence in which the reaction products form to be established for each paste.

## 2. Materials and methods

### *2.1 Materials*

HY, DHY, MgO, silica fume (SF), and sodium hexametaphosphate (SHMP, $(NaPO_3)_6$) were the starting materials used to create pastes for analysis. MasterLife SF 100, the SF source, was obtained from MasterBuilders. HY was purchased as magnesium carbonate basic from Sigma Aldrich (reagent grade, product no. 63062). DHY was synthesized by heating HY at 325 °C for 1 hour in a mechanical convection oven in air. Based on thermogravimetric data of DHY (Figure 1), 325 °C is sufficient to remove all of the structural water without causing carbonate decomposition. Reactive MgO was synthesized by calcination of HY at 700 °C for 1 hour in a muffle furnace in air, which removes all structural water, and hydroxyl and carbonate groups. To aid in the mixing of paste samples, SHMP from Sigma-Aldrich (product no. 71600) was used as a plasticizer. The XRD patterns of the raw materials (Figure S1 in Supplementary Information) show that calcination of HY caused full conversion to MgO, while a minor brucite ($Mg(OH)_2$) component is apparent in the MgO data set. This is attributed to partial rehydration of the reactive MgO powder during prolonged storage prior to the synchrotron measurements. Note that brucite was not present in the MgO powder used to synthesize paste samples since this phase is absent in all initial XRD data sets shown in Section 3.4. The SF is predominantly amorphous, with quartz and silicon carbide as minor crystalline impurities. XRD (Figure S1 in Supplementary Information) shows that the HY powder is phase pure without any contributions from impurities such as calcite.

### *2.2 Sample preparation*

Pastes were prepared by combining SF, MgO, HY, DHY, SHMP and deionized water in the amounts detailed in Table 1. More water was used for the MgO-SF-DHY paste to account for rehydration of DHY, specifically the stoichiometric amount of water required for complete reformation of HY from DHY. SHMP was added at 1 wt% of dry powder in all mixes to aid mixing, and is excluded from the theoretical wt% $CO_2$ content reported in Table 1. This table also reports the Mg/Si molar ratio of each paste, calculated from the ideal formulas for HY ($Mg_5(CO_3)_4(OH)_2 \cdot 4H_2O$), DHY ($Mg_5(CO_3)_4(OH)_2$), MgO (MgO), and SF ($SiO_2$).

Aqueous solutions were prepared first by dissolving the SHMP in water in a capped vial for 10 minutes without mixing; the solutions were then briefly manually agitated before being added to the powders. Dry powders were weighed into a separate capped vial and homogenized for 10 seconds on a vortex mixer at 2500 rpm. The solution was then added to the homogenized powder and the capped vial mixed for 1 minute on a vortex mixer at 2500 rpm. The moment of water contact is recorded as the start of the reaction. Pastes were then immediately loaded into 1.00 mm inner diameter polyimide capillaries (0.05 mm wall) and sealed on both ends with epoxy. Ex situ samples were stored at 20 °C until measurement.

**Table 1.** Mix proportions used for paste samples along with the theoretical $CO_2$ wt% of the dry powder mix and the Mg/Si molar ratio of the paste.

| Sample name | MgO (g) | SF (g) | HY (g) | DHY (g) | $H_2O$ (mL) | SHMP (g) | Mg/Si (molar ratio) | $CO_2$ (wt%*) |
|---|---|---|---|---|---|---|---|---|
| MgO-SF | 0.40 | 0.60 | 0 | 0 | 1.20 | 0.01 | 1.00 | 0 |
| MgO-SF-HY | 0.27 | 0.40 | 0.33 | 0 | 1.20 | 0.01 | 1.54 | 12.4 |
| MgO-SF-DHY | 0.27 | 0.40 | 0 | 0.33 | 1.26 | 0.01 | 1.63 | 14.7 |

* wt% relative to dry powder, excluding SHMP.

### *2.3 Data collection*

Synchrotron XRD and PDF experiments were conducted on the 28-ID-1 beamline at the National Synchrotron Light Source II at Brookhaven National Laboratory. Each sample was measured at a sample-to-detector distance of 1000 mm for XRD and 220 mm for PDF using a wavelength of 0.1665 Å. Scattering patterns were collected on a PerkinElmer XRD 1621 digital imaging plate detector; a 32 second acquisition time was used for an individual XRD image while 60 seconds was used for an individual PDF image. Each 2D image was integrated to produce a 1D data set using Dioptas v0.7.1.[36] For ex situ samples, two images were collected for each sample and averaged after integration. $LaB_6$ was also measured to provide calibration parameters. Samples were measured in 1.00 mm inner diameter polyimide capillaries (0.05 mm wall) and the capillary signal was subtracted from scattering patterns after integration. The reciprocal space total scattering data were further processed to produce the PDF data sets using PDFgetX3 v2.4.0,[37] with rpoly of 1.12 and $Q_{max}$ of 20 Å$^{-1}$.

In situ XRD and PDF data were collected for the initial 24 h of reaction for the three paste samples reported in Table 1. Table 2 shows the time of the start of the first and last XRD and PDF data collected for each of the samples. The cadence between consecutive scans was ~20 minutes with a lapse of 105 min starting at ~21 h due to loss of current in the synchrotron storage ring. There are 68 to 69 PDF and XRD images collected for each sample. For conciseness during discussion, the final scan collected for each sample will be referred to as t=24 h.

**Table 2.** In situ XRD and PDF timing.

| Sample name | First XRD scan (min) | Last XRD scan (h) | First PDF scan (min) | Last PDF scan (h) |
|---|---|---|---|---|
| MgO-SF | 60.7 | 25.0 | 48.3 | 24.7 |
| MgO-SF-HY | 56.0 | 24.9 | 45.6 | 24.7 |
| MgO-SF-DHY | 51.3 | 24.5 | 42.8 | 24.7 |

TGA was performed on the HY and DHY powders using a Perkin Elmer Pyris 1. For each sample approximately 20 mg of powder was loaded into a platinum crucible and heated from 20 °C to 900 °C at a heating rate of 10 °C/min in an $N_2$ environment.

## *2.4 Data processing*

### *2.4.1 DHY rehydration*

Representative HY/DHY peak positions were extracted from the XRD pattern of the HY powder sample and from the in situ and ex situ XRD patterns of MgO-SF-DHY and MgO-SF-HY pastes by fitting a linear background and a single Pseudo-Voigt function simultaneously to each pattern within three Q windows containing the HY (100), (110), and (011)/$(11\bar{1})$ reflections. These windows were 0.626-0.757, 0.942-1.041, and 2.107-2.239 $Å^{-1}$, respectively. The fit was implemented in the Python SciPy library;[38] the refined parameters were integrated area, center (constrained to be within its respective Q window), width, and Gaussian/Lorentzian mixing fraction, as well as background slope and intercept. Fits with $R^2 < 0.99$, evaluated after subtracting the fitted background, were rejected. This criterion excludes 12 out of 68 in situ MgO-SF-DHY fits in the (100) window (all at $t < 5$ h, before there is a well-defined peak) and 8 out of 68 in situ MgO-SF-DHY fits in the (011)/$(11\bar{1})$ window (all at $t < 3.5$ h). Fits whose refined center converged to within one measurement step (0.0034 $Å^{-1}$) of a window bound have also been rejected; this removes one further early-age MgO-SF-DHY (100) fit, also at $t < 5$ h. All remaining fits were retained. Fitted peak centers have been converted to d-spacing via $d = 2\pi/Q$.

*2.4.2 MgO dissolution and brucite/HCB formation kinetics*

To quantify MgO dissolution and brucite/HCB formation kinetics, the in situ XRD patterns were fitted over the 2.502-3.161 Å$^{-1}$ window that contains brucite/HCB ($10\bar{1}$), MgO (200), and MgO (111). To fit the patterns, all components were refined together: three Pseudo-Voigt profiles (for the three MgO and brucite/HCB peaks in this window), a linear background, and, for the carbonate-containing pastes, the as-measured powder reference patterns of HY (MgO-SF-HY) or HY and DHY (MgO-SF-DHY) as rigid components with refined scale factors. The fit was implemented in the Python SciPy library,[38] with peak areas and reference scale factors constrained to be non-negative. For the two peaks of interest, brucite/HCB ($10\bar{1}$) and MgO (200), the refined parameters were integrated area, center, width, and Gaussian/Lorentzian mixing fraction. The MgO (111) peak was found to be too weak to define its own profile as the reaction progresses and was therefore held at a fixed position, width, and Gaussian/Lorentzian mixing fraction. These fixed values were obtained by refinement using the first in situ scan for each sample where the MgO reflections are strongest; the integrated area was the only parameter refined at subsequent time steps. Fits with $R^2 < 0.99$, evaluated after subtracting the fitted background, were to be rejected; however, this criterion did not exclude any patterns. MgO (200) areas are reported normalized to the first in situ scan of each sample, and brucite/HCB ($10\bar{1}$) areas are reported as fitted.

*2.4.3 Brucite/HCB XRD peak broadness*

To assess the degree of disorder of the HCB formed in MgO-SF-HY and MgO-SF-DHY, the in situ XRD patterns for all three pastes were fitted within three Q windows containing the brucite/HCB (001), ($10\bar{2}$), and (300) reflections to quantify peak broadness. These windows were 1.172-1.462, 3.325-3.674, and 6.840-7.004 Å$^{-1}$, respectively, chosen to minimize overlap with Bragg peaks attributed to other phases. The final ten scans for each sample were fitted, corresponding to a reaction time of approximately 20 to 25 h. To fit the patterns in these windows, all components were refined together: one Pseudo-Voigt profile for the reflection of interest, a linear background, and, for the carbonate-containing pastes, the as-measured dry reference patterns of HY (MgO-SF-HY) or HY and DHY (MgO-SF-DHY) as rigid components with refined scale factors. The fit was implemented in the Python SciPy library,[38] with peak areas and reference scale factors constrained to be non-negative. For the peaks of interest, the refined parameters were integrated area, center, width, and Gaussian/Lorentzian mixing fraction, as well as background slope and intercept. The (001) window of MgO-SF-DHY was treated as an exception because the brucite/HCB peak in this window is strongly asymmetric. It was therefore refined with an independent width on each side of its center. All fits achieved $R^2 > 0.987$, evaluated after subtracting the fitted background. Full width at half maximum (FWHM) values are reported as the mean and standard deviation of the ten fits for each sample.

*2.4.4 Phase quantification from PDF data*

A least squares refinement approach was used to quantify the temporal evolution of individual phases present in the PDF data of the three samples including the formation of M-S-H. Each in situ and ex situ experimental PDF data set was fitted over 1.35-15 Å with a simulated data set consisting of potential contributions from simulated PDFs for brucite and MgO, and experimental PDFs for M-S-H, SF, water, HY (only for MgO-SF-HY samples) and DHY (only for MgO-SF-DHY samples). This range excludes the high *r* region, as the HCB in MgO-SF-DHY is too disordered to be approximated by a refined brucite structure above ~15 Å and the M-S-H pattern (digitized from Bernard et al.)[39] is not reported above 20 Å. The M-S-H reference data set was traced from the PDF of M-S-H with Mg/Si = 0.8 reported by Bernard et al.[39] The PDF data sets for SF, deionized water, HY, and DHY were the reference samples measured at the beamline.

The simulated PDFs of brucite and MgO were calculated using diffpy-cmi[40] from crystal structures obtained from the Inorganic Crystal Structure Database[41] (ICSD-79031[42] and ICSD-52026,[43] respectively). The MgO structure was refined ($a$, $\delta_2$, and $U_{iso}$ values) once per paste using the first in situ data set and the brucite structure was refined ($a$, $c$, $\delta_2$, and $U_{iso}$ values) once per paste using the final in situ data set. The refined parameters for MgO and brucite were then held constant for all other in situ and ex situ fits. All refined parameters for MgO and brucite are reported in Table S1 in the Supplementary Information. $U_{iso}$(H) for brucite was not refined and was held at 0.01 Å$^2$ due to the very weak scattering of H compared with O and Mg.

The least squares refinement consisted of refining all scale factors associated with the aforementioned phases. Prior to fitting, each paste pattern was divided by the height of the Si-O peak (~1.6 Å) so that M-S-H scale factors are proportional to the amount of total Si in the paste. The fit was implemented in the python library diffpy-cmi[40] with $Q_{damp}$ = 0.035 Å$^{-1}$ and $Q_{broad}$ = 0.013 Å$^{-1}$ (relevant for the simulated PDFs of MgO and brucite), which were obtained via a refinement of the measured $LaB_6$ standard using the python library diffpy-pdffit2.[44] All scale factors, $\delta_2$, and $U_{iso}$ values were constrained to be non-negative. $R_w$ ranged from 0.161-0.321 and is reported for each scan in Figure S8 in the Supplementary Information.

## 3. Results and discussion

### *3.1 Dehydration-induced structural changes in HY*

The TGA and derivative TGA (DTG) curves for HY and DHY are shown in Figure 1. Heating of HY to 325 °C causes dehydration via the removal of water molecules (surface and structural) but leaves the hydroxyl and carbonate groups intact. The residual masses at 900 °C (~43% for HY and ~51% for DHY) match the MgO percentages in $Mg_5(CO_3)_4(OH)_2{\cdot}4H_2O$ (43.1%) and $Mg_5(CO_3)_4(OH)_2$ (50.9%), confirming that no $CO_2$ is lost upon dehydration. The decarbonation peak is initially centered at ~520 °C in HY, which is comparable to literature sources[25,26] and shifts to ~465 °C in DHY, indicating that the loss of structural water decreases the overall crystallinity and increases the degree of disorder in the material.[45]

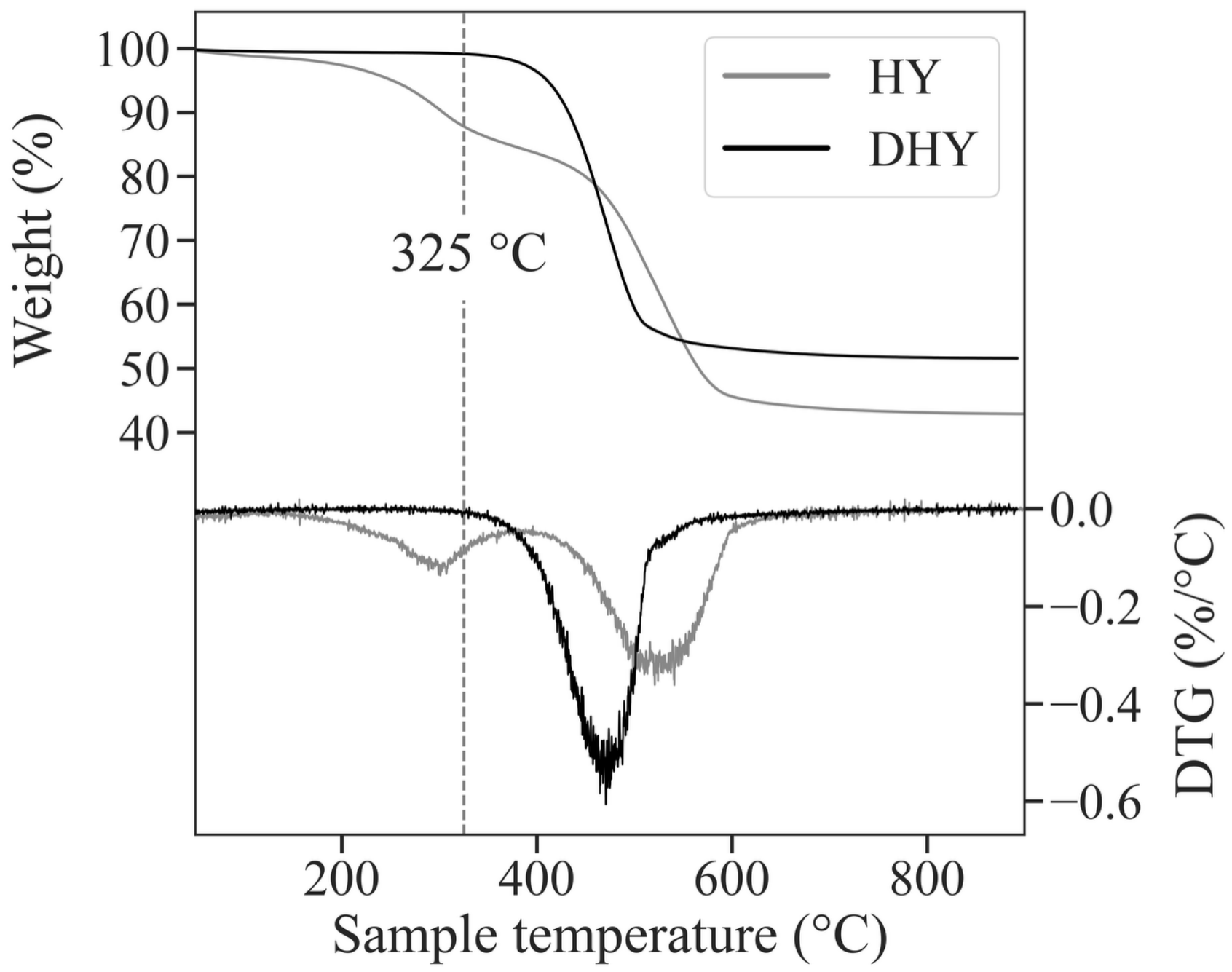

**Figure 1.** TGA (upper) and DTG (lower) curves of HY and DHY from 20-900 °C.

Comparison of the XRD patterns of HY and DHY (Figure 2) reveals several structural changes resulting from dehydration at 325 °C. Changes to the (110) and (011)/($11\bar{1}$) reflections of HY are representative of the general trends invoked by dehydration; the Bragg peaks decrease in intensity, shift to slightly larger Q values (smaller d-spacing), and in most cases are seen to broaden. The Bragg peak broadening and intensity loss in the DHY XRD pattern are consistent with increased disorder in the atomic structure.[46] An outlier to this pattern is the (100) reflection at a Q value of 0.68 $Å^{-1}$; while this Bragg peak broadens significantly (from a FWHM of 0.035 $Å^{-1}$ in HY to 0.088 $Å^{-1}$ in DHY) and loses ~67% of its integrated area, the peak location does not shift to higher Q, as would be expected if the *a* lattice parameter were to contract. Additionally, the diffuse scattering of DHY is more intense compared to HY up to ~5.3 $Å^{-1}$, further confirming the increased disorder in DHY and suggesting that it may be partially amorphous,[47] while reflections above ~5.3 $Å^{-1}$ are almost entirely absent. Finally, there is an increase in intensity at very low Q (<0.66 $Å^{-1}$), which is approaching the small-angle scattering regime. Overall, the DHY XRD pattern closely resembles that of HY heated to 280 °C reported by Yamamoto et al.[33]

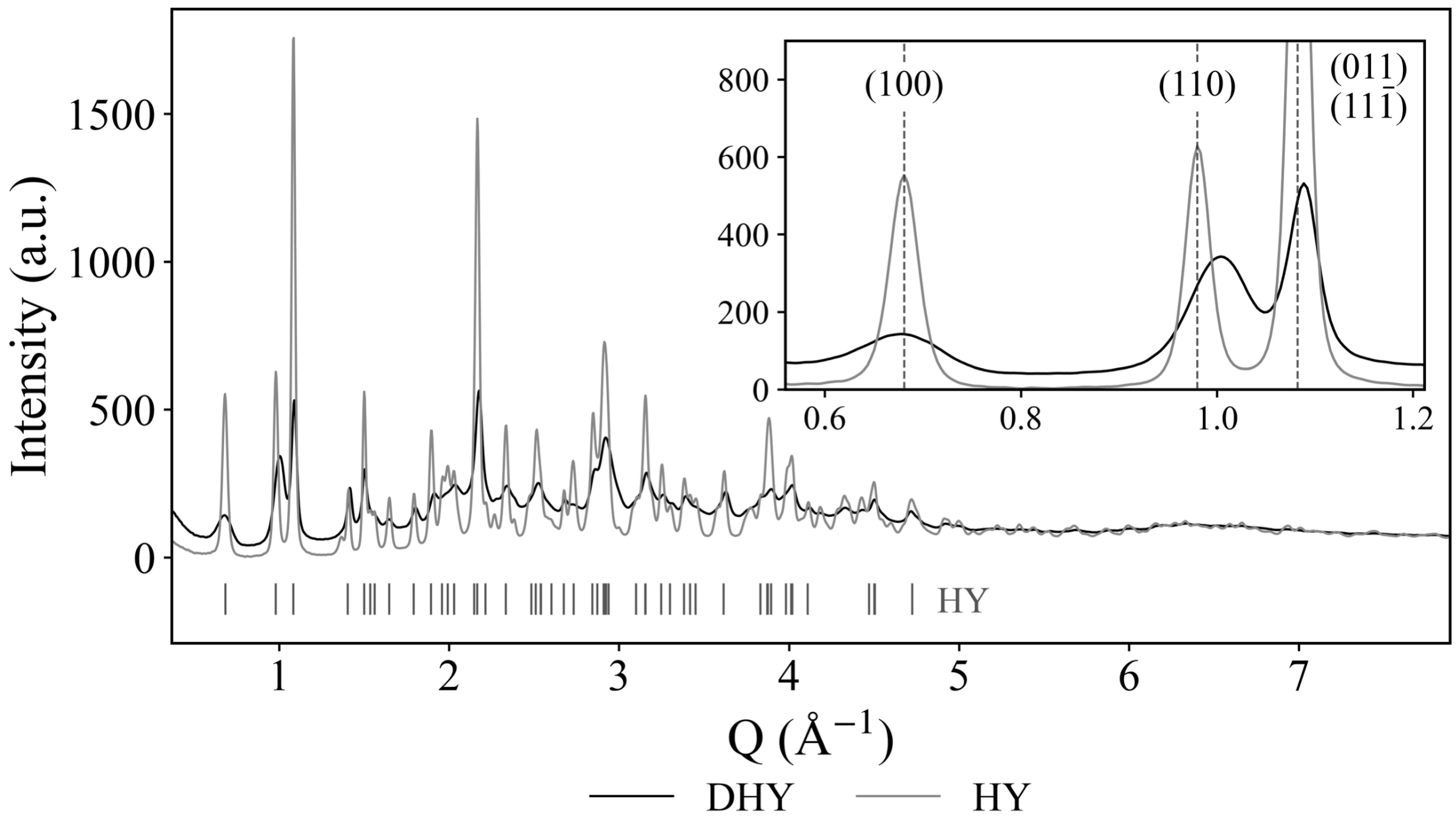

**Figure 2.** XRD patterns of HY and DHY. Inset: magnified view of (100), (110), and (011)/(11$\bar{1}$) HY reflections, showing representative general trends of Bragg peak broadening, loss of Bragg peak intensity, higher diffuse scattering intensity, and some shifts of Bragg peaks to larger Q (smaller d-spacing) in DHY. Vertical tick marks show the calculated positions of HY reflections with intensities of at least 5% of the strongest reflection.

The clear decrease in long-range order shown in the XRD data in Figure 2 is supported by PDF data in Figure 3. The PDF data show that the local atomic environment (up to ~4.5 Å) is nearly identical between HY and DHY. The short-range peaks between 1 and 4.5 Å (C-O, Mg-O, Mg-Mg, Mg-O) persist in DHY at essentially the same positions and intensities as they do in HY, demonstrating that the structural building blocks of HY are not altered by dehydration at 325 °C. Over a longer length scale (~5 to 40 Å), DHY is visibly more disordered than HY, and beyond ~40 Å it exhibits essentially no order at all.

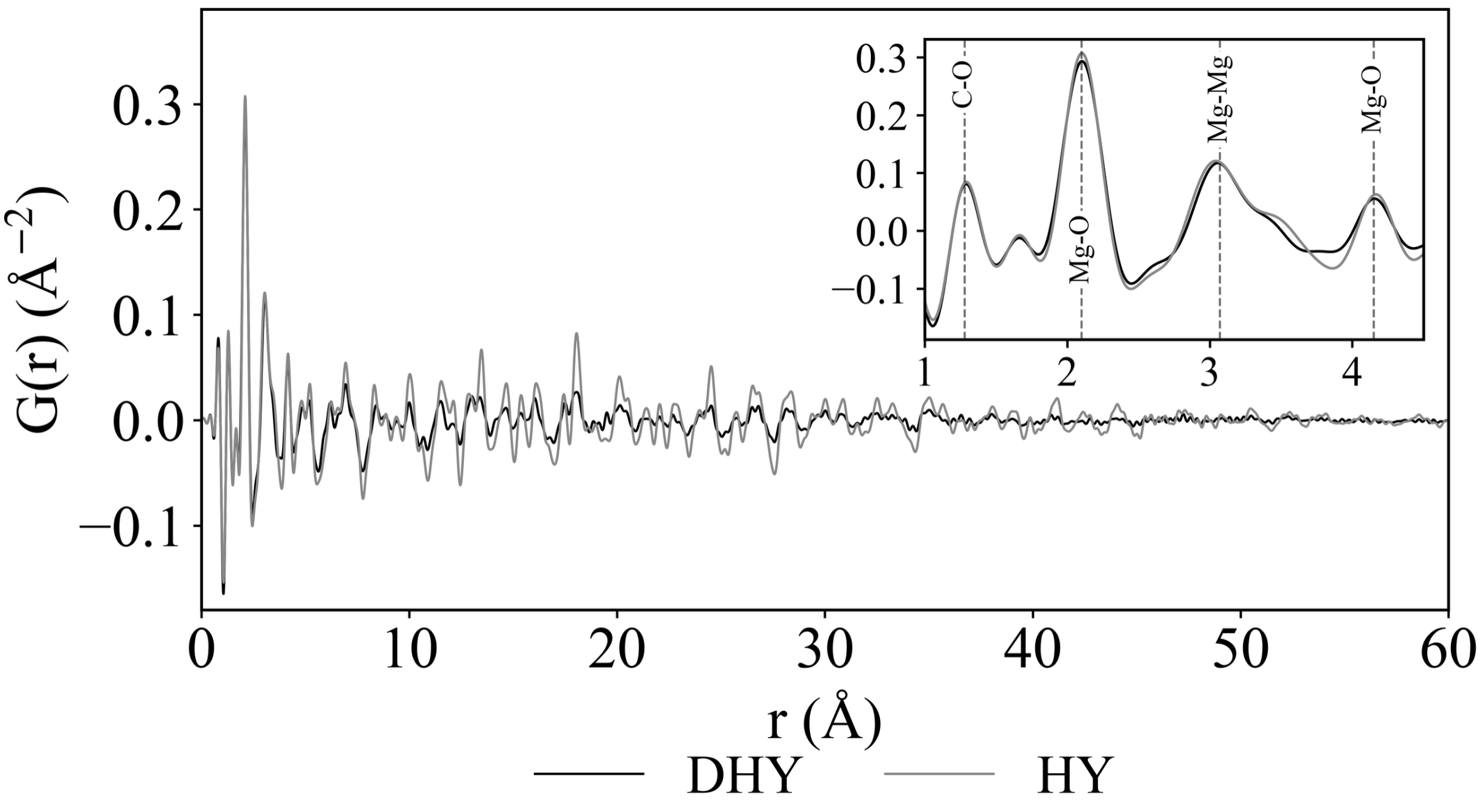

**Figure 3.** X-ray PDFs of HY and DHY. The DHY PDF is scaled to match the maximum intensity of the C-O peak at 1.28 Å in HY. Inset: magnified view of $1 < r < 4.5$ Å.

***3.2 Phase evolution during the initial 24 hours***

Figure 4 presents the in situ XRD patterns collected over the first 24 h of reaction for the three paste samples, along with identification of the crystalline phases that are present. Across all three samples, a similar sequence of events is observed: the Bragg peaks of MgO lose intensity as MgO dissolves, and brucite/HCB peaks emerge as either brucite or HCB precipitates. In the MgO-SF sample crystalline brucite is seen to precipitate, whereas in MgO-SF-HY and MgO-SF-DHY it is the more disordered HCB phase, in which (bi-)carbonate and water are incorporated into brucite-like layers[25,48] causing an increase in peak broadness. The Bragg peaks of DHY also evolve significantly over this period, where the peaks shift as DHY transforms to an HY-like phase. Quantitative insight on these processes is difficult to discern solely from the full patterns, and, as such, the following sections isolate and quantify each in turn: the evolution of the magnesium carbonate phase (Section 3.3), the kinetics of MgO dissolution and brucite/HCB formation (Section 3.4), and the degree of structural disorder of the HCB phase in the paste samples containing HY and DHY (Section 3.5).

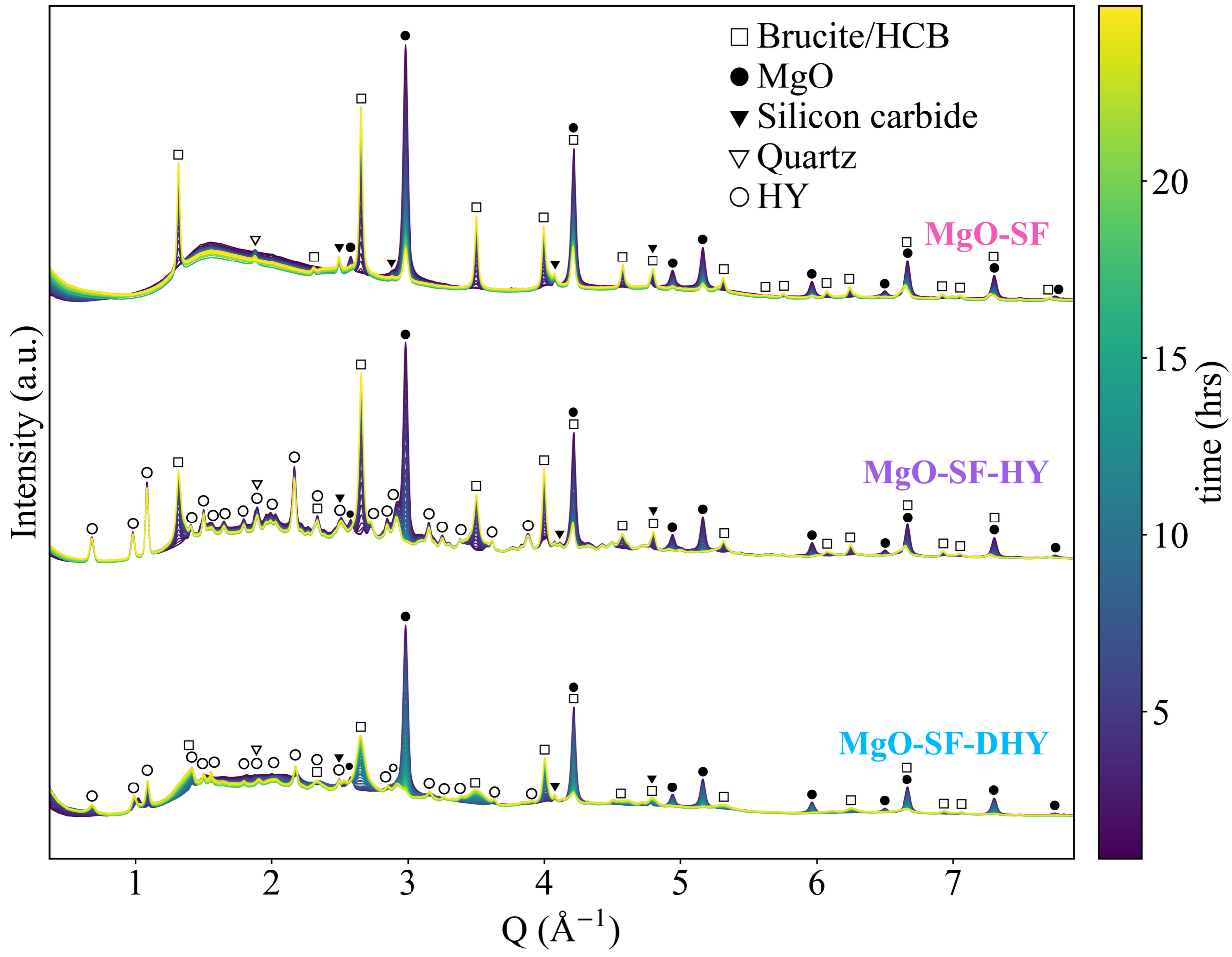

**Figure 4.** In situ XRD patterns of MgO-SF, MgO-SF-HY, and MgO-SF-DHY over the initial 24 h of reaction.

### *3.3 Evolution of DHY and HY in pastes*

Figure 5 follows the evolution of DHY reflections (100), (110), and (011)/(11$\bar{1}$) over the full course of the measured reaction in the MgO-SF-DHY sample, combining the in situ data (~1 to 24 h) with the ex situ measurements at 1, 3, 7, 14, and 42 days. The evolution of these Bragg peaks gives insight into how DHY participates in the early reaction period compared to HY in the MgO-SF-HY system (Figure 6). Early in the reaction (less than ~5 h), the (110) and $(011)/(11\bar{1})$ reflections are weak and broad, and are located at higher Q values than the HY reference position. For comparison, these same HY reflections ((100), (110), $(011)/(11\bar{1})$) are plotted in Figure 6 for MgO-SF-HY. The locations of these three HY reflections in the MgO-SF-HY sample are effectively identical across all time points.

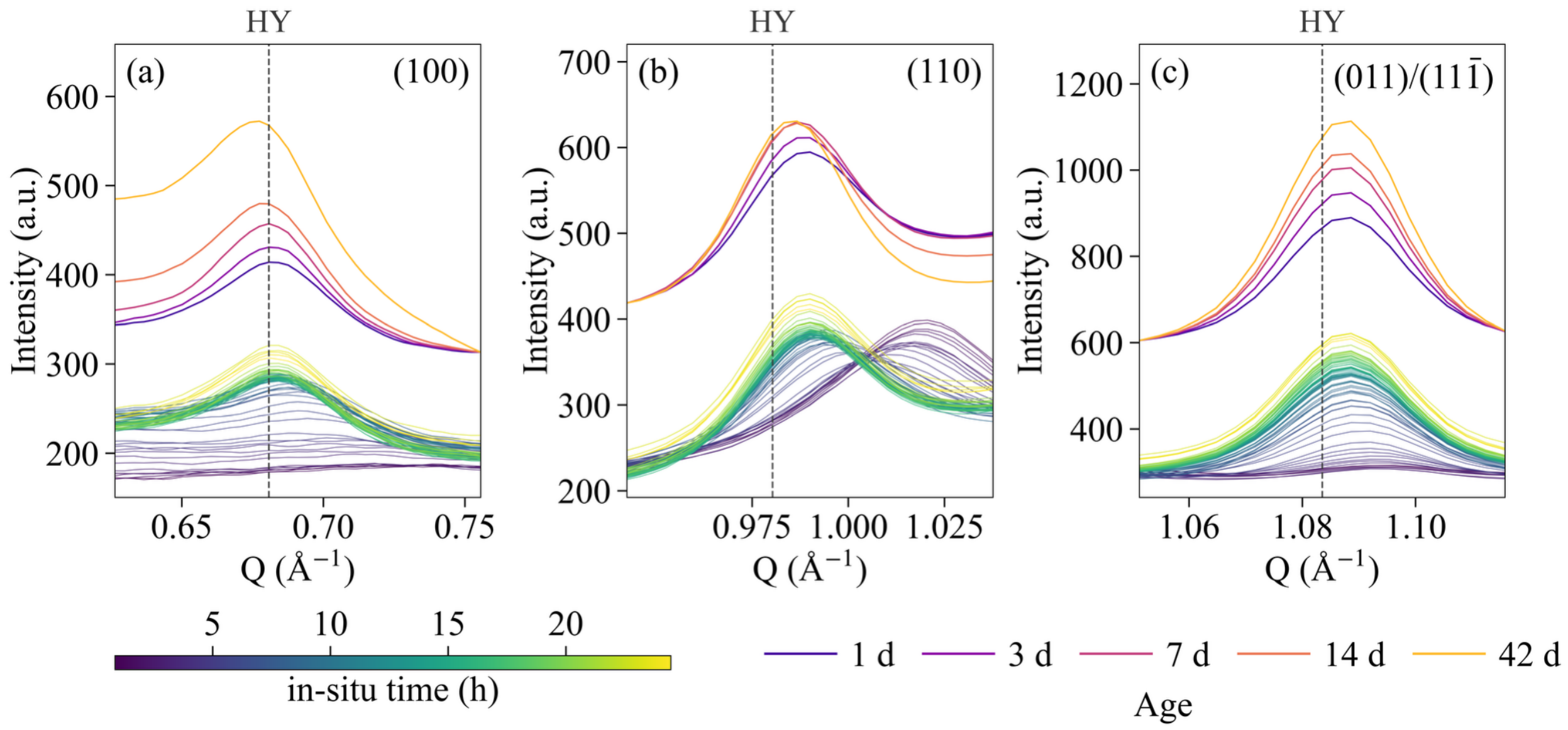

**Figure 5.** XRD patterns showing the DHY/HY reflections (a) (100), (b) (110), and (c) $(011)/(11\bar{1})$, during the course of the reaction (in situ and ex situ) for MgO-SF-DHY.

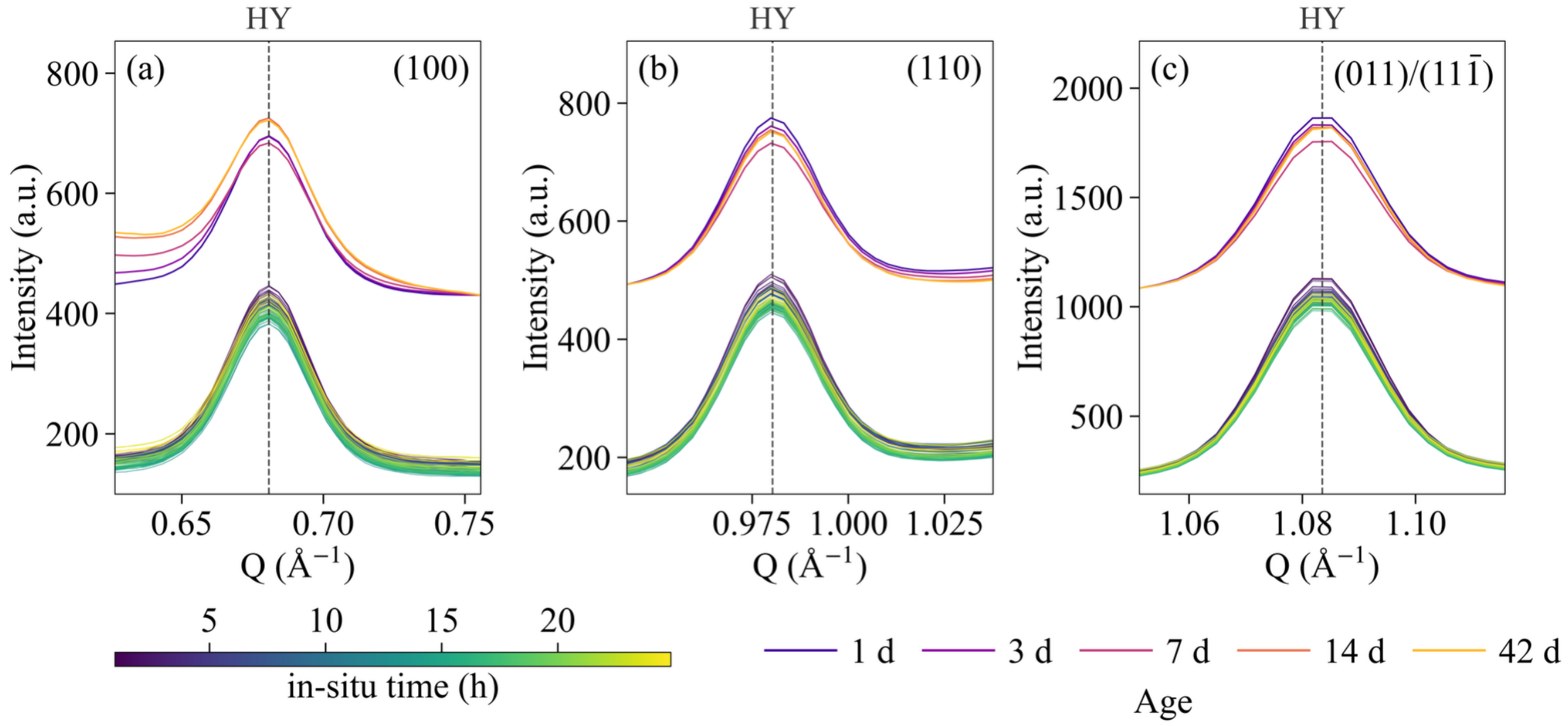

**Figure 6.** XRD patterns showing the HY reflections (a) (100), (b) (110), and (c) (011)/(11$\bar{1}$), during the course of the reaction (in situ and ex situ) for MgO-SF-HY.

Figure 7 tracks the d-spacing of the (100), (110), and (011)/(11$\bar{1}$) HY reflections in both MgO-SF-HY and MgO-SF-DHY based on XRD peak position, providing a clearer picture of the behavior of the magnesium carbonate phase in MgO-SF-DHY during the early hours of reaction. For the (110) reflection (Figure 7b), the d-spacing at the beginning of the reaction is 6.16 Å and steadily increases toward the HY value of 6.41 Å over the initial ~10 h, after which it makes only incremental progress toward the HY value over 42 days. The (011)/(11$\bar{1}$) reflection (Figure 7c) follows the same trend, increasing from 5.75 Å toward the HY value of 5.80 Å. Similar to the (110) reflection, this Bragg peak shift behavior shows that the majority of the structural reformation of HY in MgO-SF-DHY happens within the initial ~10 hours. By 42 days, both the (110) and (011)/(11$\bar{1}$) planes in MgO-SF-DHY have not quite fully recovered the spacing of the HY reference. The (100) reflection (Figure 7a) is seen to behave somewhat differently to the (110) and (011)/(11$\bar{1}$) reflections. First, this reflection is absent in the MgO-SF-DHY sample until ~5 h, after which the d-spacing starts at 9.01 Å before increasing to the HY value of 9.23 Å around 7 days. By 42 days the d-spacing of the (100) reflection in the MgO-SF-DHY sample is actually larger than that of pure HY by 0.04 Å. For reference, Figure 7 confirms that the d-spacings of the (100), (110), and (011)/(11$\bar{1}$) HY reflections in MgO-SF-HY are constant through the entire reaction period, varying by no more than 0.006, 0.004, and 0.002 Å, respectively, across all in situ and ex situ measurements.

Figure 8 shows the time evolution of the FWHM of these same Bragg peaks. In MgO-SF-HY, the widths remain essentially constant at ~0.034 $Å^{-1}$ for (100), ~0.028 $Å^{-1}$ for (110), and ~0.024 $Å^{-1}$ for (011)/(11$\bar{1}$). In MgO-SF-DHY, the (110) peak narrows sharply over the first ~10 h, over the same period as the rapid d-spacing increase in Figure 7, while the (100) and (011)/(11$\bar{1}$) peaks in MgO-SF-DHY narrow more gradually. None of the three peaks in MgO-SF-DHY reach the FWHM values in MgO-SF-HY, even after 42 days. Sample factors that affect peak widths in a crystalline phase are small crystallite size, nonuniform strain, and/or stacking faults,[47] so the larger FWHM values indicate that the structural order of the magnesium carbonate phase in MgO-SF-DHY is only partially recovered. Overall, these data show that the DHY phase undergoes rehydration in the MgO-SF-DHY paste, where water molecules enter the atomic structure of DHY leading to an expansion of the unit cell (Figure 7) and a partial recovery of crystallinity (Figure 8).

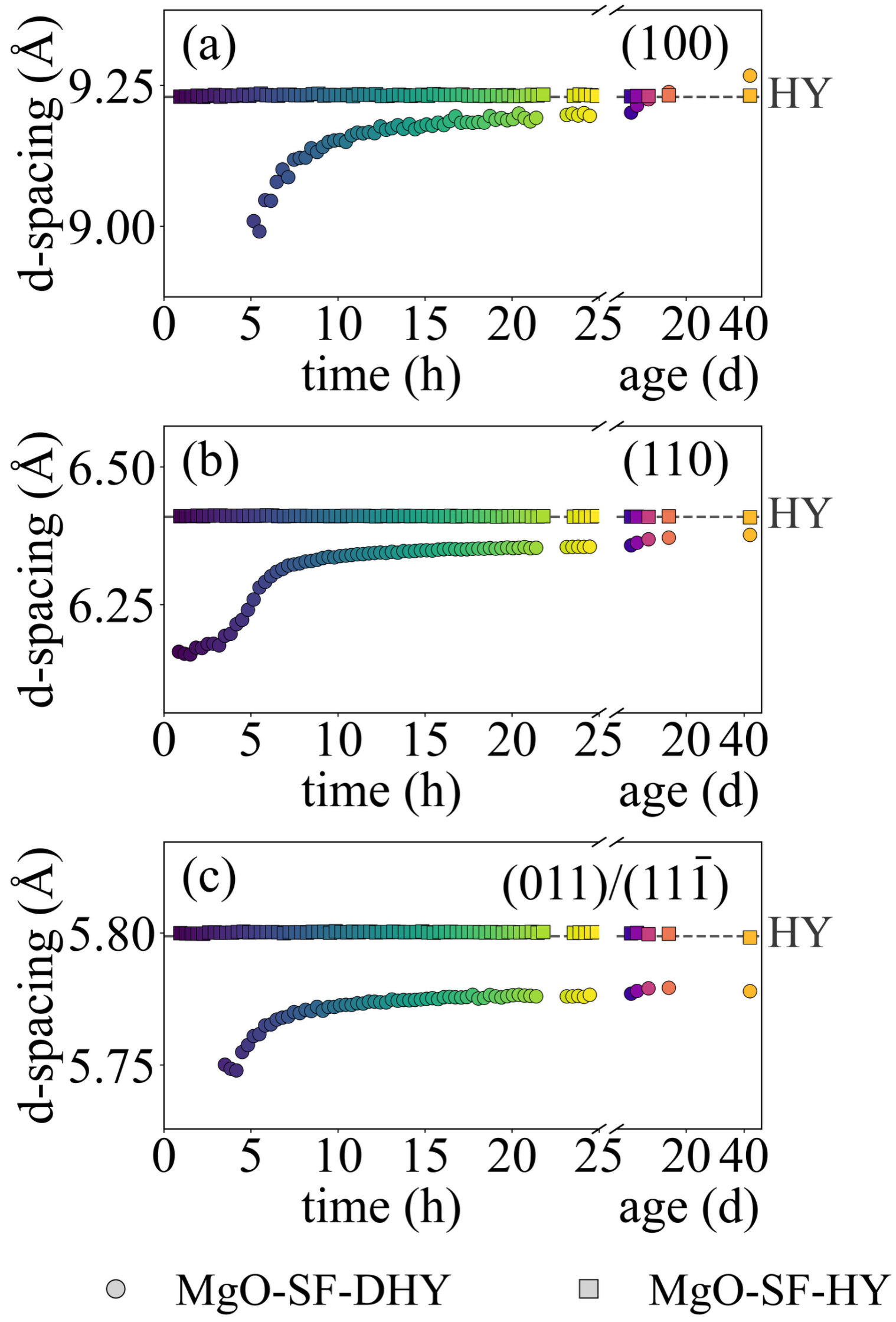

**Figure 7.** d-spacing from XRD patterns for HY reflections (100) (a), (110) (b), and (011)/(11$\bar{1}$) (c) during the course of the reaction (in situ and ex situ) in MgO-SF-HY and MgO-SF-DHY.

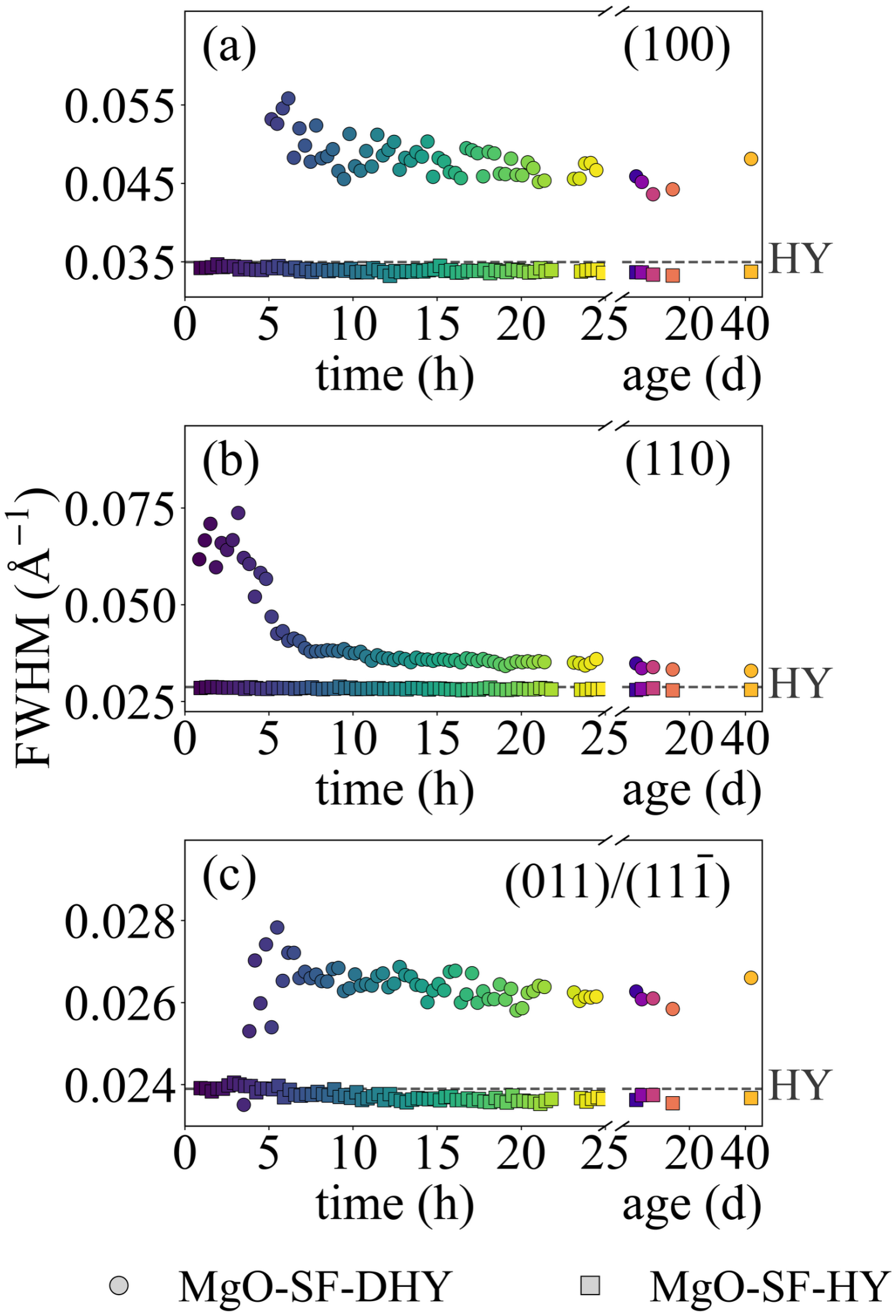

**Figure 8.** FWHM of Bragg peaks from XRD patterns for HY reflections (100) (a), (110) (b), and (011)/(11$\bar{1}$) (c) during the course of the reaction (in situ and ex situ) in MgO-SF-HY and MgO-SF-DHY.

### *3.4 MgO dissolution and brucite/HCB formation kinetics during the initial 24 hours*

Figure 9 shows the in situ XRD data collected over the first 24 h of reaction for the three paste samples, focused on the 2.50-3.16 Å$^{-1}$ region that contains contributions from MgO and brucite/HCB. The HY dry powder is shown for reference in MgO-SF-HY (Figure 9b), and both the HY and DHY dry powders are shown in MgO-SF-DHY (Figure 9c). All three paste samples follow the same broad pattern over the first 24 h, during which the MgO (200) reflection decreases in intensity as MgO dissolves and the brucite/HCB ($10\bar{1}$) reflection grows due to brucite/HCB precipitation. In the neat paste (i.e., MgO-SF) the precipitating phase is crystalline brucite, whereas in the MgO-SF-HY and MgO-SF-DHY pastes it is the more disordered HCB (particularly for MgO-SF-DHY), in which (bi-)carbonate and water are incorporated into brucite-like layers.[48]

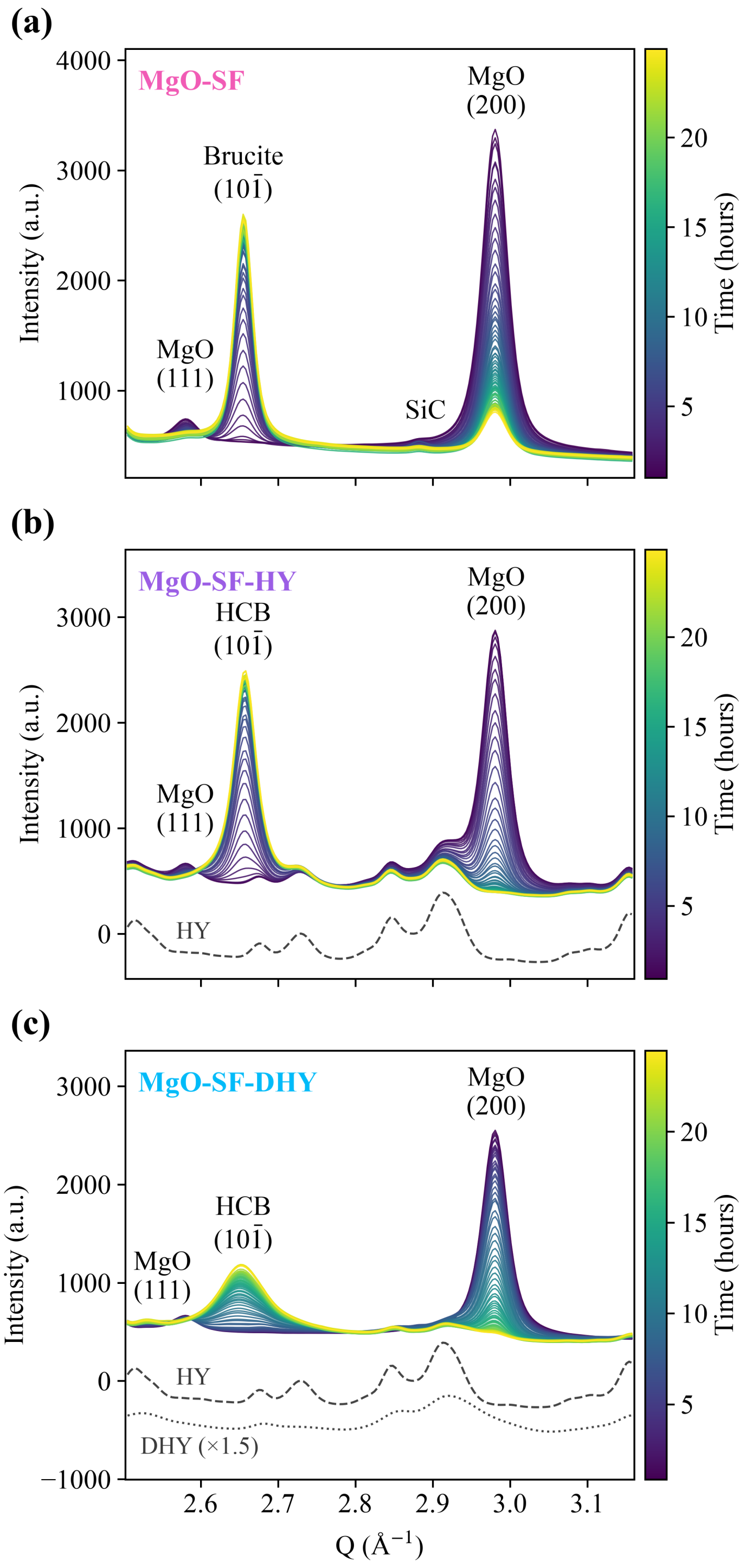

**Figure 9.** In situ XRD patterns showing MgO (200) dissolution and brucite/HCB (10$\bar{1}$) formation during the initial 24 h for (a) MgO-SF, (b) MgO-SF-HY, and (c) MgO-SF-DHY.

Figure 10 quantifies this behavior by plotting the extent of MgO dissolution and formation of brucite/HCB over the initial 24 h. The extent of MgO dissolution (Figure 10a), α, was calculated using the area of the (200) reflection at time *t* relative to the initial area in the first in situ scan, i.e., $\alpha = 1 - (A_t/A_0)$. Note that since α is calculated relative to the first in situ scan rather than to the time of mixing, α of 0 does not mean that no MgO dissolved before the first in situ measurement at ~1 h (Table 2). Any MgO dissolution before the first in situ scan was not captured, and so α is a lower bound on the extent of reaction. Formation of brucite/HCB (Figure 10b) was tracked by growth of the $(10\overline{1})$ reflection peak area. Raw peak area units are not comparable between samples, so the curves are compared using the time at which each process begins, the time at which it is half complete (α of 0.5 for MgO dissolution, half of the 24 h plateau for brucite/HCB peak area), and the time at which it reaches a plateau.

In MgO-SF, MgO dissolution begins from the first in situ scan, whereas brucite is not detected until ~2 h. MgO dissolution and brucite precipitation are half complete at ~5.5 h and ~4.5 h, respectively. Brucite peak areas reach a plateau at ~8-10 h, whereas MgO dissolution continues slowly for the remainder of the in situ period and does not reach completion by 24 h. At the end of the in situ data, α is 0.88, meaning that ~12% of MgO remains. This is confirmed in Figure 9a, where a small but well-defined MgO (200) reflection is apparent at the end of the in situ measurement. Beyond the in situ window, MgO remains detectable only in the ex situ MgO-SF samples, where it persists through 7 days and is fully consumed only by 14 days (Figure 13). Note that although a minor brucite component was detected in the XRD pattern of the reference MgO sample (Figure S1 in Supplementary Information), there are no brucite Bragg peaks in the initial in situ MgO-SF pattern, indicating that the brucite present in the MgO pattern was confined to that particular capillary.

In MgO-SF-HY, both processes begin at similar times to those in MgO-SF; MgO dissolution begins from the first in situ scan and HCB first appears at ~2 h. In this sample, MgO dissolution and HCB precipitation are both half complete at ~4.5-5 h. Unlike in MgO-SF, MgO dissolution tracks HCB growth, with both leveling off at ~10 h and α reaching 0.99-1.00 by the end of the in situ data, which is confirmed by the absence of MgO peaks in the final in situ patterns (Figure 9b) and in all ex situ patterns (Figure 13). MgO-SF-DHY is kinetically distinct from the other two pastes. MgO dissolution begins from the first in situ scan but proceeds slowly at first, reaching half completion at ~8.5 h, ~3-4 h later than in the other two pastes. Nonetheless, it reaches α of 0.98 by the end of the in situ data, which is confirmed by the absence of MgO peaks in the final in situ patterns (Figure 9c) and in all ex situ patterns (Figure 13). HCB growth in this paste is delayed even further, with no HCB detected until ~6.5 h, ~4.5 h later than in the other two pastes. The HCB peak area reaches half of its plateau at ~10.5 h, ~6 h later than in the other two pastes. The absence of HCB in the first ~6.5 h coincides with the rapid rehydration of DHY to an HY-like phase that reaches a maximum rate at ~5 h and is largely complete by ~10 h (Section 3.3), suggesting that the early reaction period in this sample is dominated by the evolution of the HY/DHY phase before substantial HCB formation begins. Figure 10 also shows that ~1/3 of MgO has dissolved prior to any formation of HCB in the MgO-SF-DHY sample. The fate of this dissolved magnesium is investigated via PDF analysis in Section 3.6.

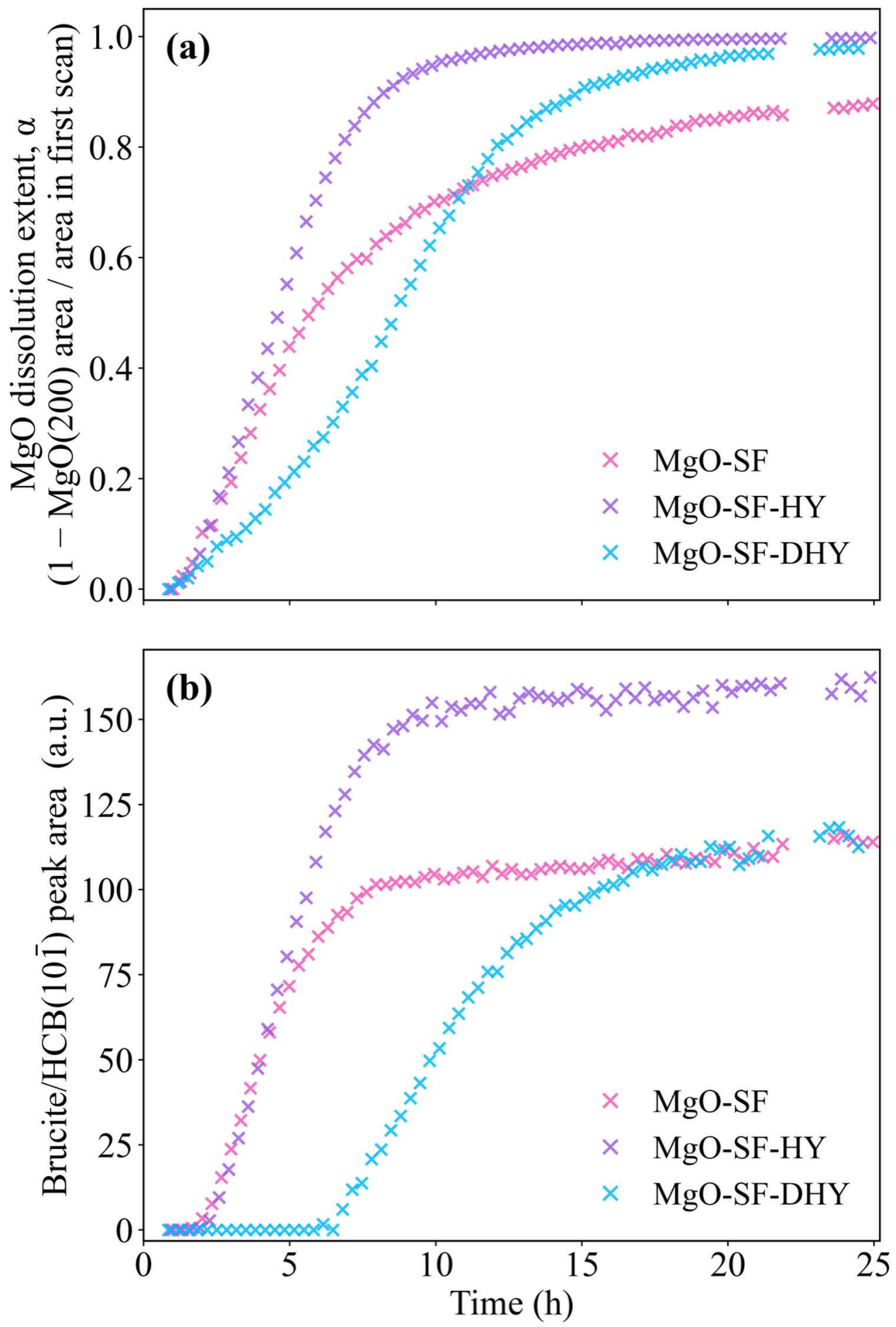

**Figure 10.** (a) MgO dissolution extent, α, and (b) brucite/HCB (10$\bar{1}$) peak area, determined from in situ XRD patterns for all three paste samples during the initial 24 h of reaction. $\boldsymbol{\alpha = 1 - (A_t/A_0)}$, where $\boldsymbol{A_t}$ and $\boldsymbol{A_0}$ are the areas of the MgO (200) reflection at time *t* and in the first in situ scan, respectively. α is calculated relative to the first in situ scan (~1 h, Table 2), so any MgO dissolution before the first scan is not captured. The brucite/HCB peak area is given in arbitrary units.

### *3.5 Character of HCB formed in MgO-SF-HY and MgO-SF-DHY*

Figure 11 examines the (001), $(10\bar{2})$, and (300) reflections of brucite/HCB in the three pastes over the course of the in situ measurement, providing insight on the character of the HCB phase in the MgO-SF-HY and MgO-SF-DHY samples. Moreover, Figure 12 reports the fitted FWHM for each of these reflections averaged over the final 10 in situ scans (~20-25 h), when the brucite/HCB peaks are most prominent. Figure 11 clearly shows the difference in the degree of disorder between the HCB that forms in MgO-SF-HY and MgO-SF-DHY, where the HCB reflections in the MgO-SF-HY sample are structurally more similar to the crystalline brucite reflections seen in the MgO-SF sample than those of HCB in MgO-SF-DHY. The brucite/HCB reflections associated with the stacking-direction component (i.e., (001) and $(10\bar{2})$) are seen to significantly broaden for the DHY-containing sample. The (001) reflection broadens from a FWHM of 0.0242 ± 0.0001 $Å^{-1}$ in MgO-SF to 0.0413 ± 0.0001 $Å^{-1}$ in MgO-SF-HY and 0.1338 ± 0.0017 $Å^{-1}$ in MgO-SF-DHY. The $(10\bar{2})$ reflection follows a similar trend, from 0.0290 ± 0.0001 $Å^{-1}$ in MgO-SF to 0.0444 ± 0.0001 $Å^{-1}$ in MgO-SF-HY and 0.1158 ± 0.0008 $Å^{-1}$ in MgO-SF-DHY. These observations generally agree with the literature description of HCB that water and (bi-)carbonate are incorporated in the interlayer stacking direction in varying amounts.[28,48] In contrast to the clear broadening of the (001) and $(10\bar{2})$ reflections, the in-plane (300) reflection broadens only slightly due to the addition of the carbonate phase (i.e., HY or DHY), from 0.0277 ± 0.0003 $Å^{-1}$ in MgO-SF to 0.0310 ± 0.0005 $Å^{-1}$ in MgO-SF-HY and 0.0419 ± 0.0016 $Å^{-1}$ in MgO-SF-DHY. However, there is a shift of the (300) peak to larger Q (smaller d-spacing) with the addition of carbonate, signifying an in-plane contraction as (bi-)carbonate and water are incorporated into the overall structure. Hence, the behavior of the (300) FWHM also suggests that the brucite-like layers themselves are modified in MgO-SF-DHY, but to a far smaller degree than the interlayer.

Of the three samples, MgO-SF-DHY forms by far the most disordered HCB. The reflections are so weak and broad that the MgO-SF-DHY patterns are scaled by a factor of two in Figure 11 to make the peak shapes visible. The (001) reflection is also broad and strongly asymmetric, with the low Q side (larger d-spacings) approximately 1.9 times wider than the high Q side, consistent with a distribution of interlayer spacings rather than a single well-defined value.[29] Additionally, unlike in these data, in lab-based XRD investigations this HCB peak is sometimes resolved into two separate components.[29,48] Since the MgO-SF-DHY mixture contains only slightly more $CO_2$ than MgO-SF-HY (12.4 wt% of dry powders in MgO-SF-HY vs. 14.7 wt% in MgO-SF-DHY), the significant difference in HCB disorder is likely not attributable to this compositional difference alone and instead may reflect the greater reactivity of the carbonate phase in the DHY-containing system. A comparison of paste systems containing HY and DHY with equal amounts of $CO_2$ is necessary to elucidate the underlying effects of dehydrating the HY phase on the character of HCB formed in these paste systems and is an objective for future work.

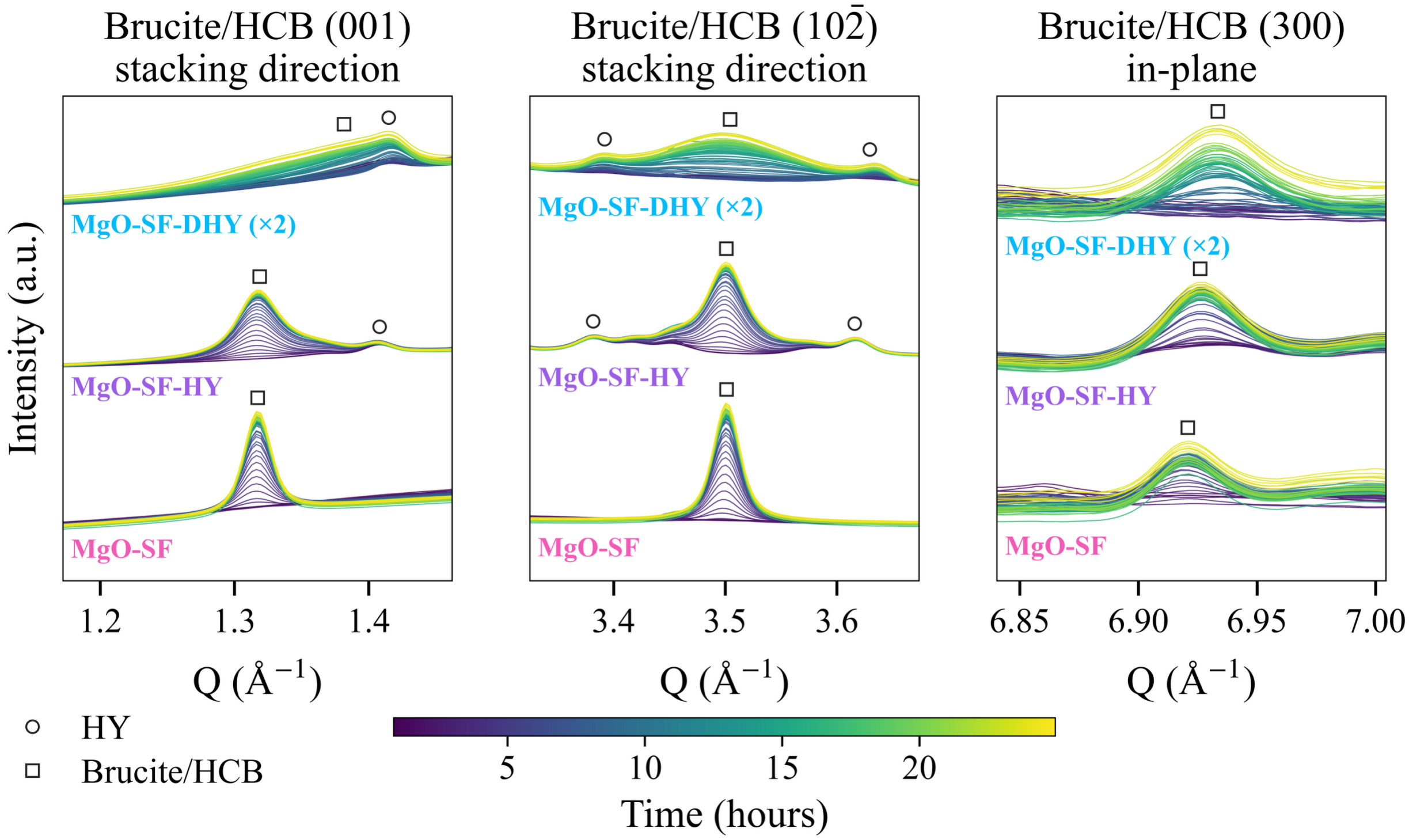

**Figure 11.** In situ XRD patterns for three representative brucite/HCB peaks in the MgO-SF, MgO-SF-HY, and MgO-SF-DHY samples. MgO-SF-DHY patterns are scaled by ×2 to more clearly see the peak shape.

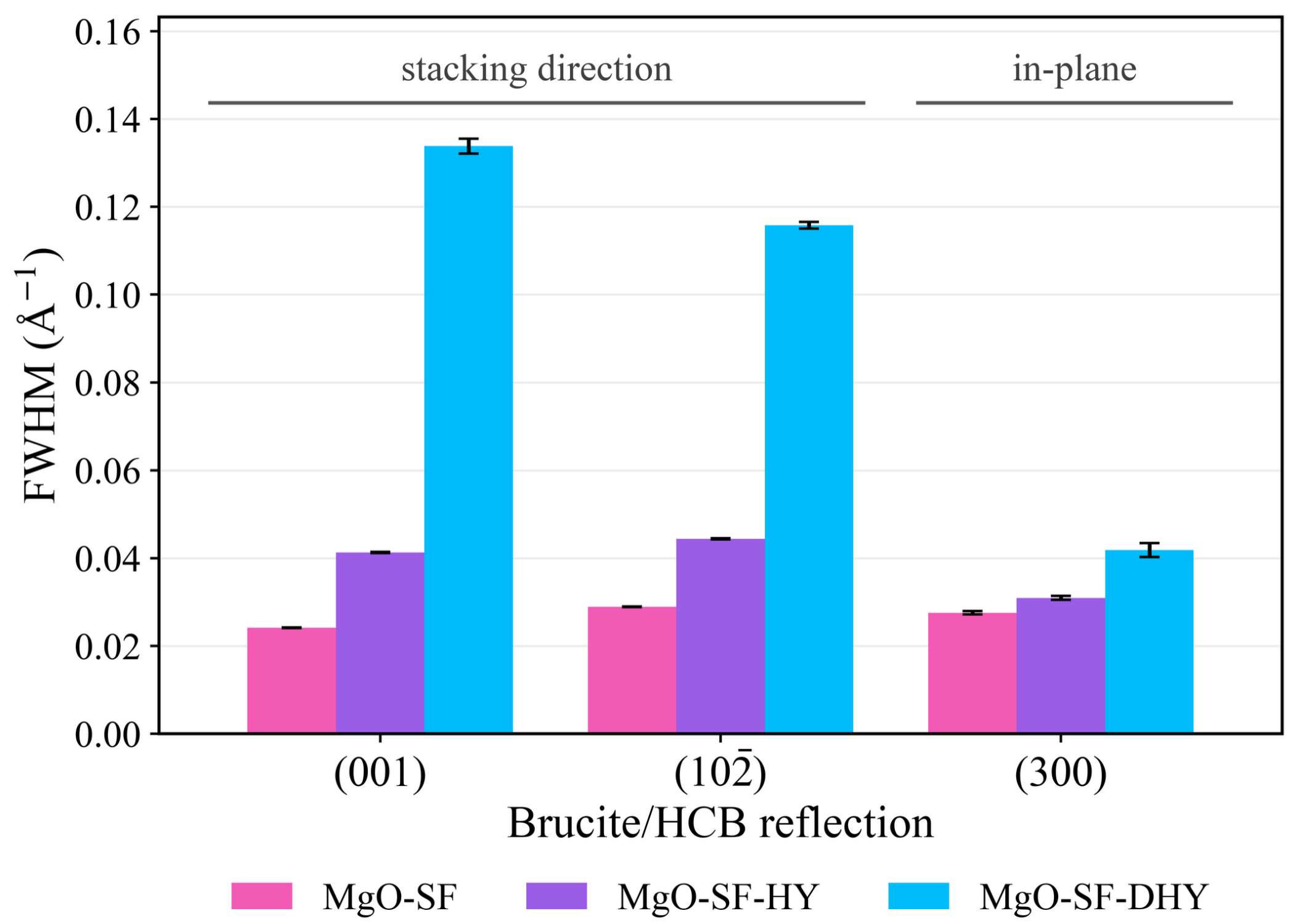

**Figure 12.** FWHM as an average of the final ten in situ XRD patterns (~20-25 h) for brucite/HCB reflections in the stacking direction (001 and 10$\bar{2}$) and in-plane direction (300). Error bars represent the standard deviation.

### *3.6 M-S-H formation*

Figure 13 shows the ex situ XRD patterns of the three paste samples at 1, 3, 7, 14, and 42 days, allowing the crystalline phases in each binder to be identified and followed over time. Since each ex situ measurement time consisted of a separate capillary measured independently, absolute peak intensities are not directly comparable between ages or between samples due to potential differences in paste densities. The patterns are therefore interpreted in terms of which phases are present, rather than their absolute intensities. The XRD patterns of all three samples exhibit little change in the crystalline phases that are present between 1 and 42 days, in contrast to the rapid evolution over the first 24 h (Sections 3.2-3.5). In particular, the MgO-SF sample is dominated by sharp brucite Bragg peaks between 1 and 42 days while MgO-SF-HY is dominated by HY and HCB peaks. MgO-SF-DHY largely consists of the reformed HY Bragg peaks and HCB contributions that are fairly stable through 14 days, after which the HCB contributions are seen to decrease.

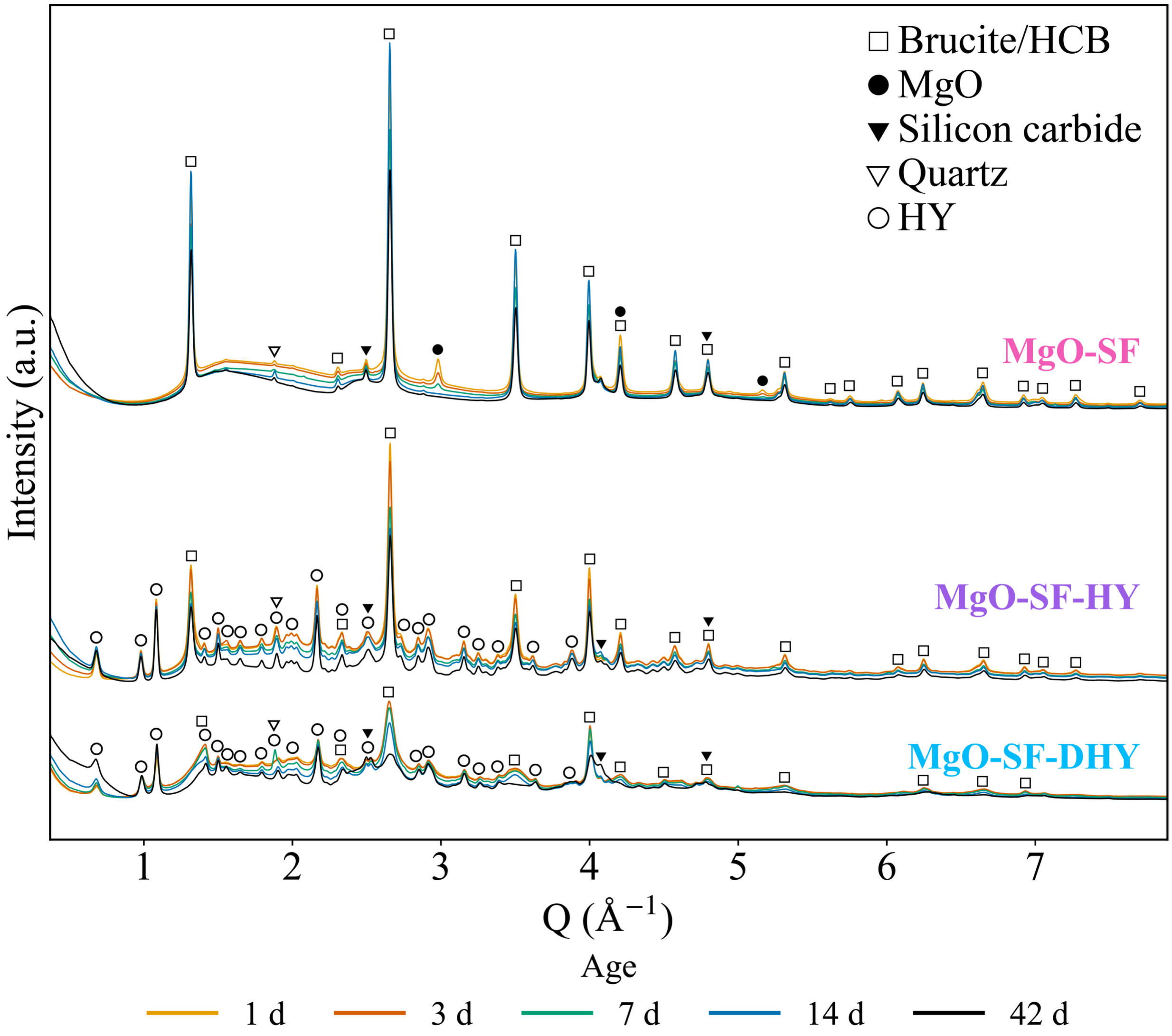

**Figure 13.** Ex situ XRD patterns over a Q range of 0.37 to 7.89 Å$^{-1}$ for the three paste samples at 1, 3, 7, 14, and 42 days.

The most notable change in the ex situ XRD patterns is the emergence of M-S-H alongside the disappearance of HCB peaks in the MgO-SF-DHY sample at 42 days. Figure 14 a-c shows XRD patterns in the 3.88-4.18 Å$^{-1}$ region for the three pastes, highlighting the region where diffuse scattering associated with M-S-H is expected should this phase emerge (shaded grey). In MgO-SF and MgO-SF-HY this region shows no appreciable growth in diffuse scattering with age, even at 42 days. In contrast, MgO-SF-DHY has a broad feature that develops in this region at 42 days, suggesting M-S-H formation. Figure 14 d-f shows the same paste samples in the 3.36-3.69 $Å^{-1}$ region to highlight the behavior of the brucite/HCB ($10\bar{2}$) reflection over the same time period. While a direct comparison of ex situ peak intensities between samples is not possible (i.e., comparing relative HCB peak intensities), the complete collapse of this HCB reflection in the MgO-SF-DHY paste suggests that the consumption of HCB and the formation of M-S-H are tied in this system. Moreover, M-S-H is seen to exist in the paste system containing the more disordered form of HCB (Section 3.5), i.e., MgO-SF-DHY. In MgO-$SiO_2$ binder systems, it is generally thought that brucite/HCB reacts with dissolved silica to form M-S-H, so a less stable HCB will convert to M-S-H more rapidly provided sufficient dissolved silica is available for M-S-H formation.[11,27]

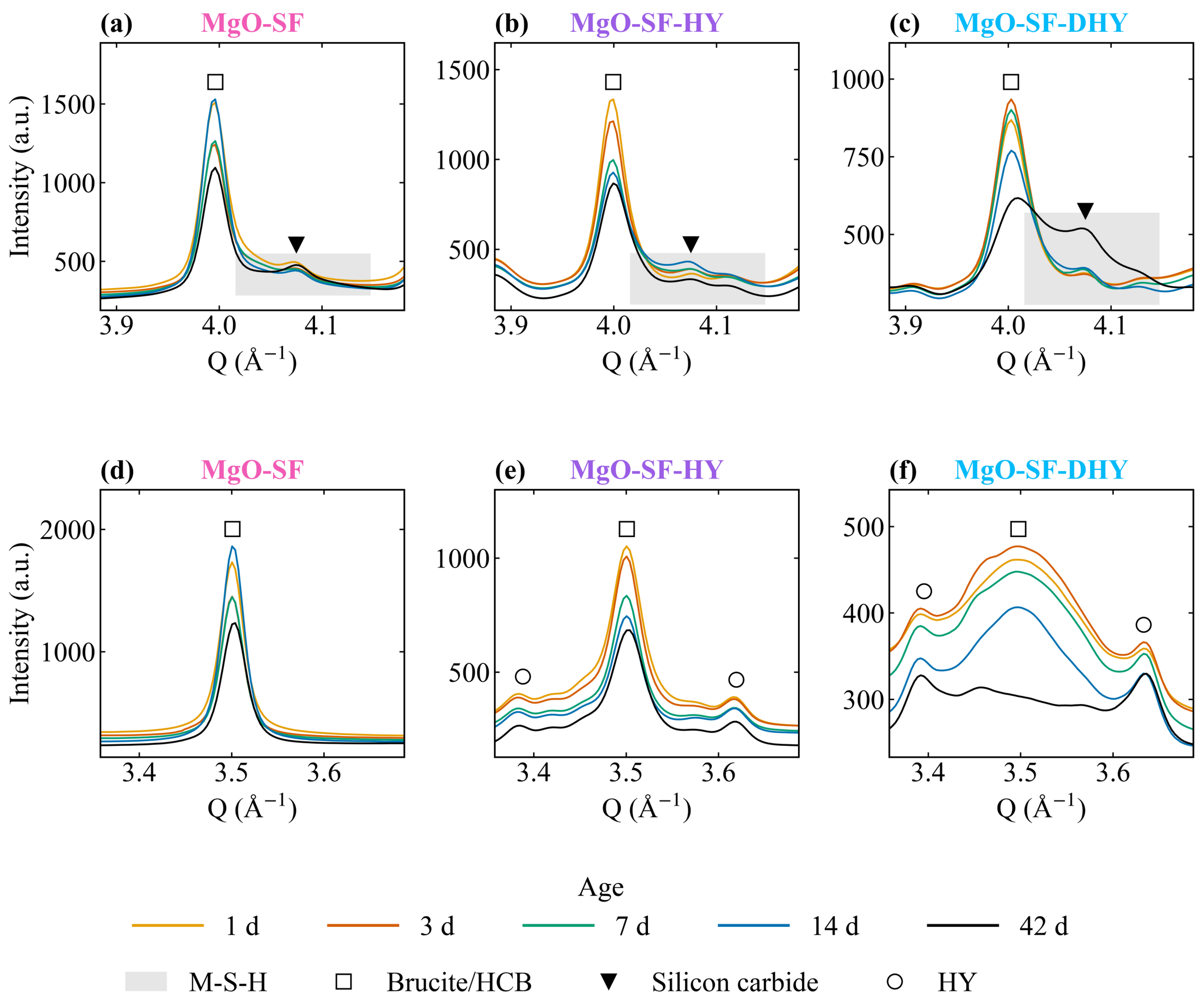

**Figure 14.** Ex situ XRD patterns from 1-42 days for the three paste samples focused on the Q range 3.88 to 4.18 Å$^{-1}$ (a-c for MgO-SF, MgO-SF-HY, and MgO-SF-DHY, respectively), with a grey rectangle highlighting the region where M-S-H is expected to appear, and Q range of 3.36 to 3.69 Å$^{-1}$ (d-f for MgO-SF, MgO-SF-HY, and MgO-SF-DHY, respectively), a region with a prominent brucite/HCB reflection. M-S-H formation is seen to occur in the MgO-SF-DHY sample.

While the XRD data show clear M-S-H formation only in MgO-SF-DHY at 42 days, PDF analysis is sensitive to detecting smaller quantities of M-S-H that may be difficult to fully resolve in XRD patterns. M-S-H is a poorly crystalline phase with coherence length of 1.2 nm,[39] so its formation should be visible in the mid-range region of the PDF. Therefore, the in situ and ex situ PDF data for all three pastes were fitted from 1.35-15 Å using a least squares approach with phase pure PDFs of M-S-H, brucite, MgO, water, SF, and additionally HY or DHY for MgO-SF-HY or MgO-SF-DHY, respectively, as outlined in Section 2.4.4. The PDF for each phase in this range is shown in Figure S5 in the Supplementary Information. Representative PDF fits along with residuals are shown in Figure S6 (in situ PDF data) and Figure S7 (ex situ PDF data) in the Supplementary Information. Prior to fitting, each PDF data set was normalized to the intensity of the Si-O peak, so the resulting M-S-H scale factor (Figure 16) is proportional to the amount of M-S-H per total Si and therefore can be compared between pastes made with different amounts of Si. The scale factors for all phases and the $R_w$ of each fit are shown in Figure S8 in the Supplementary Information. Figure 15 shows the corresponding ex situ PDF data from 1.35-15 Å alongside the M-S-H reference PDF (in situ PDF data from 1-40 Å and 1-10 Å are displayed in Figures S2 and S3, respectively, and ex situ PDF data from 1-40 Å are displayed in Figure S4 in the Supplementary Information), which shows the growth of prominent M-S-H correlations in all three pastes, most notably in MgO-SF-DHY between 14 and 42 days at ~5.4 and ~8.2 Å.

It is clear from Figure 16 that M-S-H begins to form immediately from the start of the in situ data (~45 min) in all three pastes. After the onset, the shapes of the M-S-H growth curves are similar to those of the MgO dissolution (from XRD data, Figure 10a, and PDF data, Figure S8c in Supplementary Information), in which the reaction is fastest in the first ~10 hours and then slows toward the end of the in situ period. Mirroring the trend of MgO dissolution, M-S-H formation in the in situ data reaches a plateau first in MgO-SF-HY, followed by MgO-SF-DHY and then MgO-SF. By 24 h, all three pastes converge to a similar M-S-H content per Si. To our knowledge this is the first time M-S-H has been directly observed to precipitate in situ in an MgO-$SiO_2$ paste within the initial 24 h of hydration. In pastes made from MgO and SF with practical water contents, M-S-H formation is typically resolved in XRD patterns between one week and three months,[7,27] although M-S-H has been detected as early as 1 day in pastes with more water (w/s = 2)[13] and in dilute suspensions using fumed silica in place of SF (w/s = 45).[11]

The identification of M-S-H formation on the order of hours within these pastes has implications for the origins of early-age strength development in MgO-$SiO_2$ binders. M-S-H is considered to be the main strength-giving phase in MgO-$SiO_2$ binders,[5,17] but since these systems gain significant compressive strength before M-S-H is resolved by XRD, the early-age strength gains have been previously attributed to brucite (which generates only low compressive strength)[26,28,49] or to particle packing.[16] This PDF analysis shows M-S-H formation within the first day in all three pastes, and the amount of M-S-H formed in the neat paste (i.e., MgO-SF) at the end of the in situ measurement (~24 h) is roughly half of the M-S-H present at 42 days. From 1 to 42 days, all three pastes continue to form M-S-H (Figure 16). Up to 14 days, MgO-SF-HY shows the largest M-S-H growth (1.5× MgO-SF at 14 days), yet between 14 and 42 days, MgO-SF-DHY nearly doubles in M-S-H content, ending with a scale factor of 0.32, 1.5× that of MgO-SF-HY and 1.8× that of MgO-SF. This is most visible in the PDF as peak growth at ~5.4 and ~8.2 Å (Figure 15). This increase coincides with a ~40% decrease of the fitted brucite/HCB scale factor in this sample (Figure S8b in Supplementary Information), tying the consumption of this disordered HCB phase to M-S-H formation, in agreement with the XRD data (Figure 14).

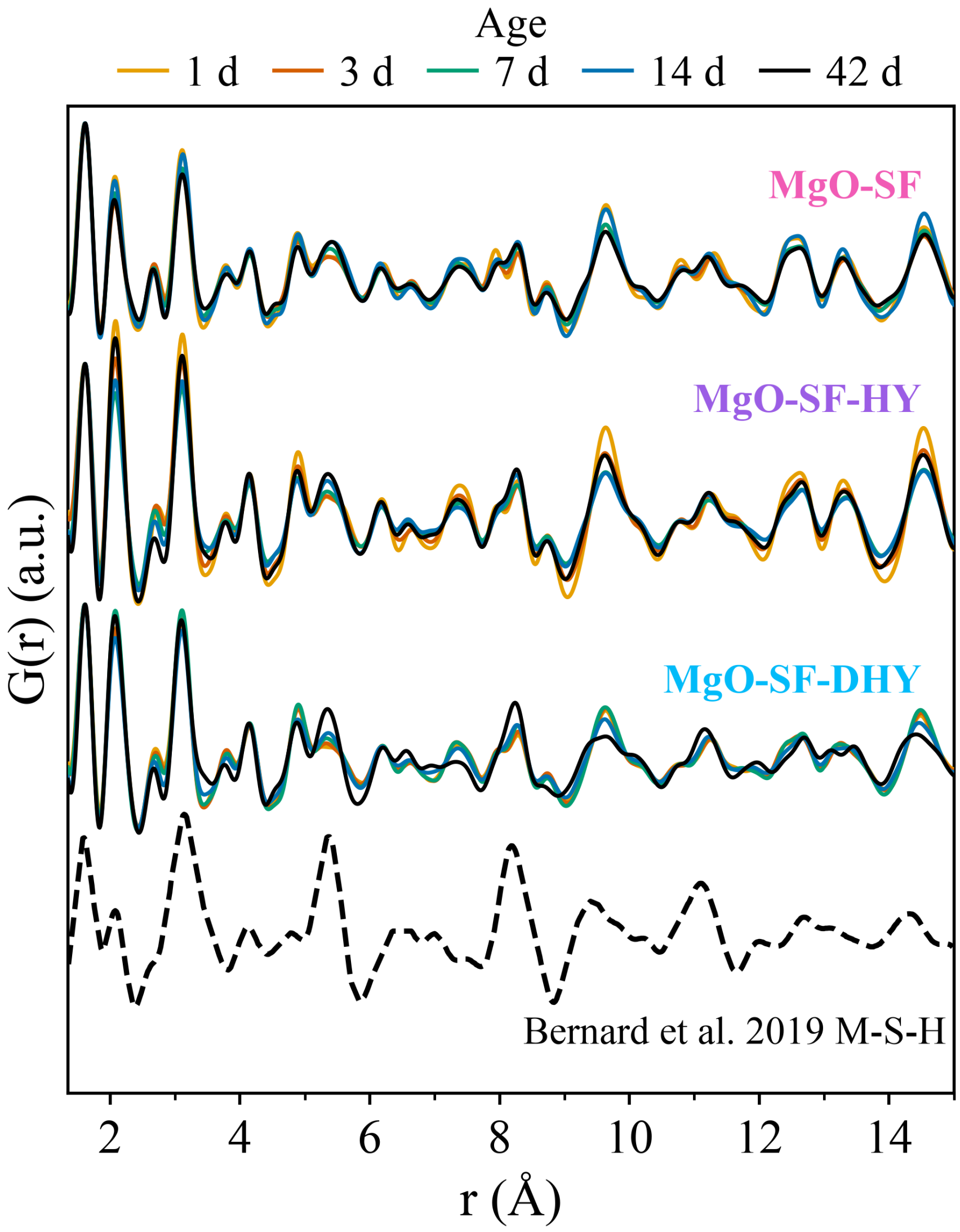

**Figure 15.** Ex situ PDFs from 1-42 days of the three paste samples, normalized to the Si-O peak at ~1.6 Å, alongside a published M-S-H pattern[39] from 1.35-15 Å.

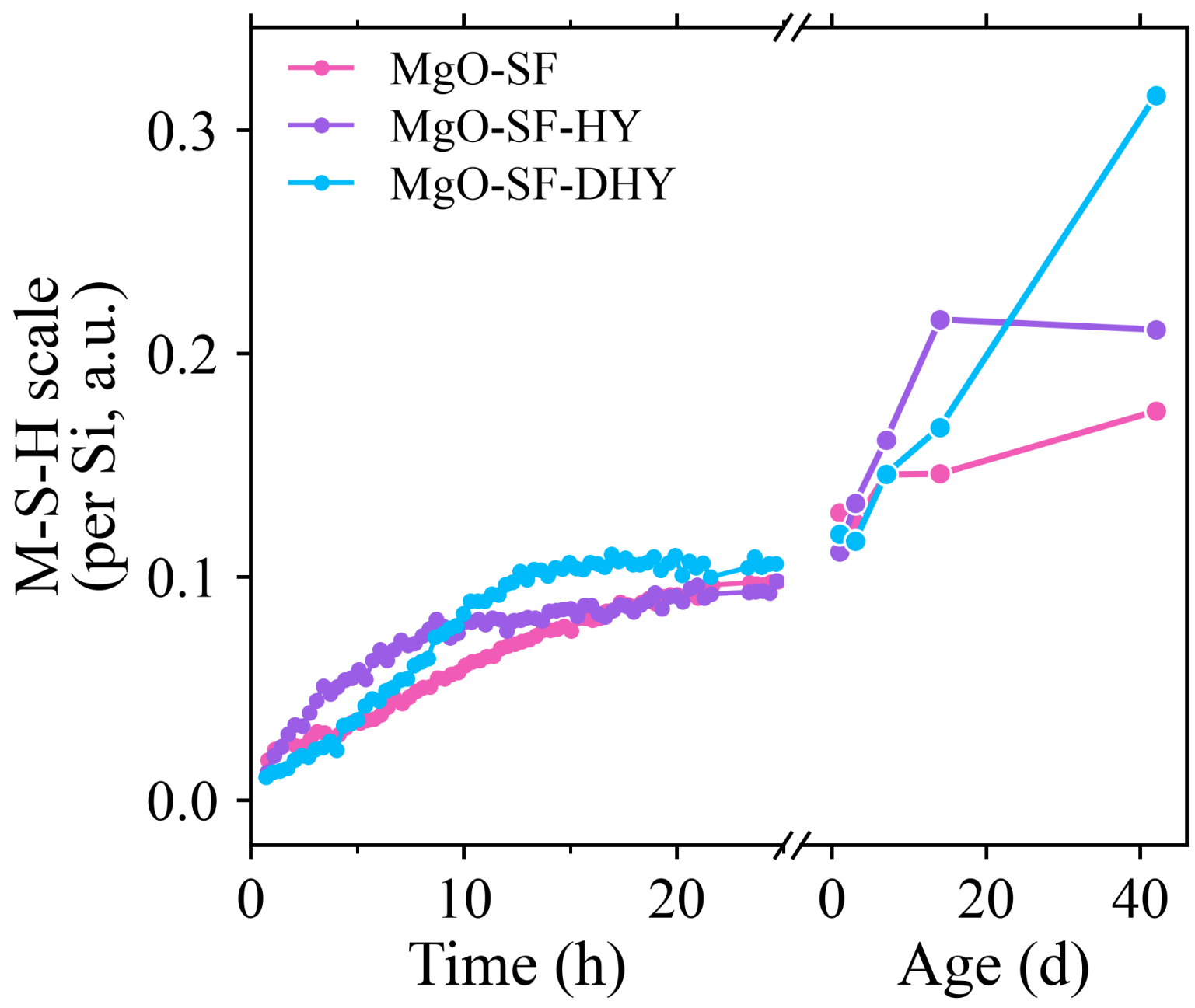

**Figure 16.** M-S-H scale factor for the three paste samples obtained from the least squares refinement of the PDF data.

Combining the in situ XRD and PDF results shows that M-S-H formation can occur in the absence of brucite or HCB during the early reaction period. In all three pastes, M-S-H is present in the first in situ PDF scan (~45 min, Figure 16), when the MgO and SF scale factors are already decreasing (Figure S8 in Supplementary Information), whereas brucite or HCB appears only later: after ~2-3 h in MgO-SF and MgO-SF-HY (~2 h from XRD, ~3 h from PDF) and after ~6-8 h in MgO-SF-DHY (~6.5 h from XRD, ~7-8 h from PDF). Therefore, the magnesium in the earliest M-S-H comes from the dissolving MgO rather than from brucite or HCB. This does not contradict the established role of brucite and HCB as metastable precursors to M-S-H in MgO-$SiO_2$ binders,[16,18,27] but rather shows that direct precipitation of M-S-H from dissolving MgO and $SiO_2$ occurs from the outset and that the precursor pathway becomes dominant only at later ages, as seen in MgO-SF-DHY between 14 and 42 days. The clearest example of the direct pathway is in MgO-SF-DHY, where XRD shows that ~1/3 of the MgO dissolves before HCB appears (Section 3.4), and PDF reveals that this magnesium is used to form M-S-H (Figure 16). The MgO-SF and MgO-SF-HY samples also show evidence of M-S-H forming via the direct pathway during the initial ~2 h of reaction, at which point ~1/10 of the MgO present in the first scan has reacted.

## 4. Conclusions

In this investigation, we used in situ and ex situ synchrotron X-ray diffraction (XRD) and pair distribution function (PDF) analysis to characterize how the addition of hydromagnesite (HY) or dehydrated hydromagnesite (DHY) to an MgO-$SiO_2$ binder alters the early reaction kinetics and the characteristics of the hydration products. Three pastes were investigated: MgO-SF, MgO-SF-HY, and MgO-SF-DHY, where silica fume (SF) was used as the silica source. In situ measurements were performed out to ~24 h, while ex situ measurements were conducted at 1, 3, 7, 14, and 42 days. From these experiments it was found that when HY is added to the binder, a slightly disordered hydrous carbonate-containing brucite (HCB) forms in place of crystalline brucite, and the rate and extent of MgO dissolution are enhanced, reaching complete dissolution within the first 24 h. Addition of DHY is seen to suppress the kinetics of MgO dissolution from the outset, HCB formation is delayed by ~4.5 h, and DHY is seen to rapidly rehydrate to an HY-like phase prior to the emergence of HCB. This suggests that the early reaction period in this sample is dominated by the evolution of the magnesium carbonate phase (i.e., DHY) before substantial HCB formation can begin. However, in both carbonate-containing binders (i.e., MgO-SF-HY and MgO-SF-DHY), MgO is fully dissolved by 24 h, compared to >7 days in the neat binder. The MgO-SF-DHY binder forms a far more disordered HCB than that of MgO-SF-HY, a difference we attribute to the disorder and higher reactivity of DHY compared with HY rather than the slightly higher $CO_2$ content of DHY.

Insight into M-S-H formation was provided by fitting the PDF data over 1.35-15 Å, revealing that M-S-H is present in all three pastes within one hour of mixing, before any brucite or HCB forms. The earliest M-S-H therefore precipitates directly from the dissolving MgO and $SiO_2$ rather than from a brucite or HCB precursor. The clearest case is MgO-SF-DHY, where ~1/3 of the MgO dissolves before any HCB appears, and M-S-H forms throughout this period. All three pastes reach similar amounts of M-S-H per Si at the end of the in situ data. Notably, in the MgO-SF paste, the amount of M-S-H formed at 24 h is roughly half of what forms by 42 days, even though XRD did not resolve M-S-H in this paste at any age. This early-age detection of M-S-H in all of the pastes suggests that M-S-H contributes to the early-age strength gain in MgO-$SiO_2$ binders. At 42 days, XRD and PDF analysis agree that MgO-SF-DHY contains more M-S-H than MgO-SF and MgO-SF-HY, and both techniques tie this growth between 14 and 42 days to the consumption of HCB. Future efforts should compare HY- and DHY-containing pastes with equal $CO_2$ contents to separate the effect of disorder in the magnesium carbonate phase from that of the carbonate dosage.

**Supplementary Information**

XRD patterns of MgO, HY, and SF; In situ and ex situ X-ray PDFs of MgO-SF, MgO-SF-HY, and MgO-SF-DHY; Reference PDFs of individual phases; Real space phase quantification results for MgO-SF, MgO-SF-HY, and MgO-SF-DHY; PDF refined parameters for MgO and brucite

**Author contributions**

**Jordan Mravca-Bailey**: Conceptualization, Methodology, Investigation, Visualization, Writing - Original Draft. **Claire E. White**: Methodology, Writing - Review & Editing, Supervision, Funding acquisition

**Conflicts of interest**

The authors declare that they have no known competing financial interests or personal relationships that could have appeared to influence the work reported in this paper.

**Acknowledgments**

This work was financially supported by funding from the Andlinger Center for Energy and the Environment at Princeton University. This research used beamline 28-ID-1 of the National Synchrotron Light Source II, a U.S. Department of Energy (DOE) Office of Science User

Facility operated for the DOE Office of Science by Brookhaven National Laboratory under Contract No. DE-SC0012704.

# Early-age formation of MgO-$SiO_2$ binders and the effect of hydromagnesite and dehydrated hydromagnesite: An in situ synchrotron X-ray diffraction and pair distribution function study

Jordan Mravca-Bailey[a,b], Claire E. White[a,b*]

[a] Department of Civil and Environmental Engineering, Princeton University, USA

[b] Andlinger Center for Energy and the Environment, Princeton University, USA

[*]Corresponding author.

## Supplementary Information

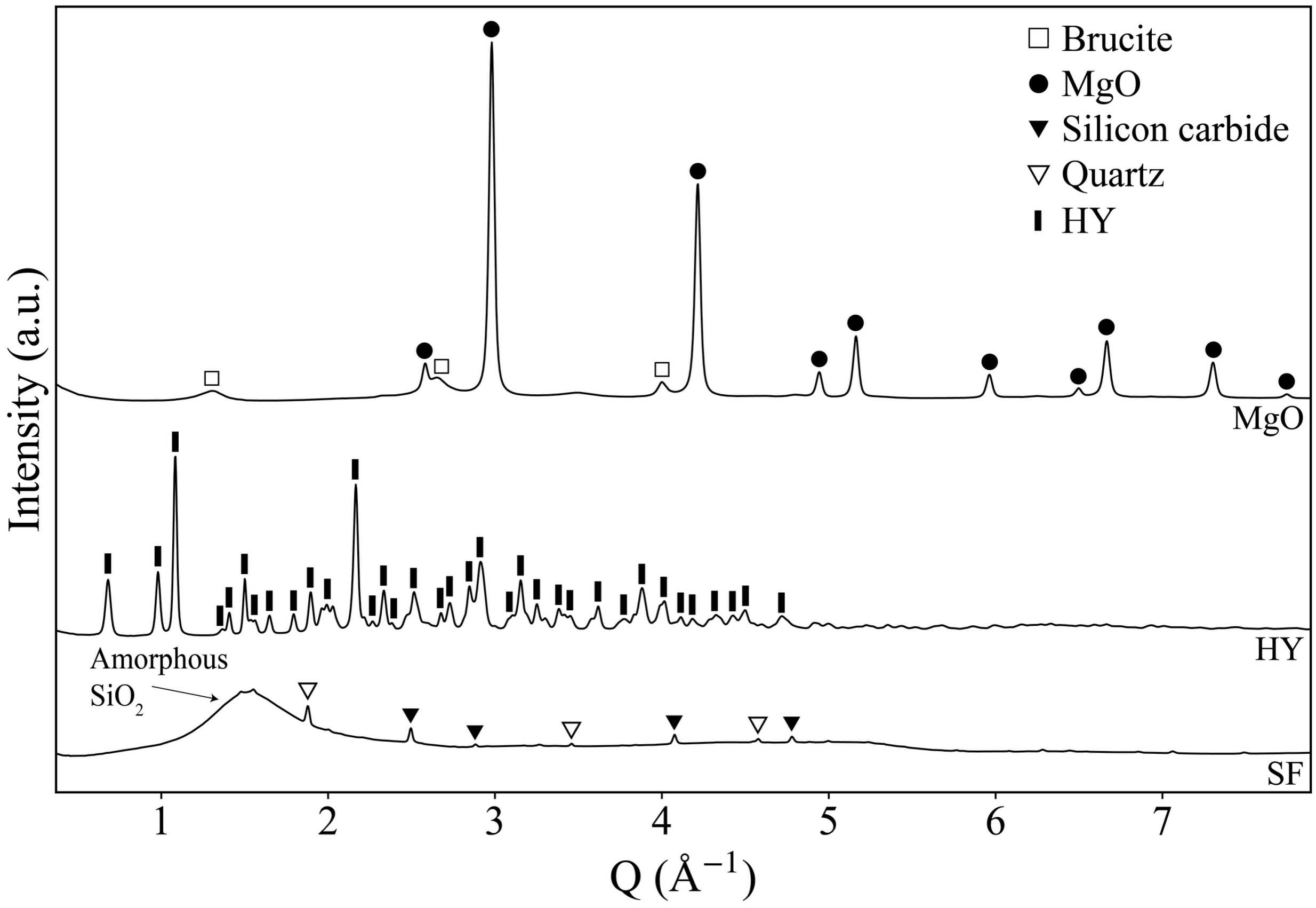

**Figure S1.** XRD patterns of MgO, HY, and SF used to synthesize the paste samples.

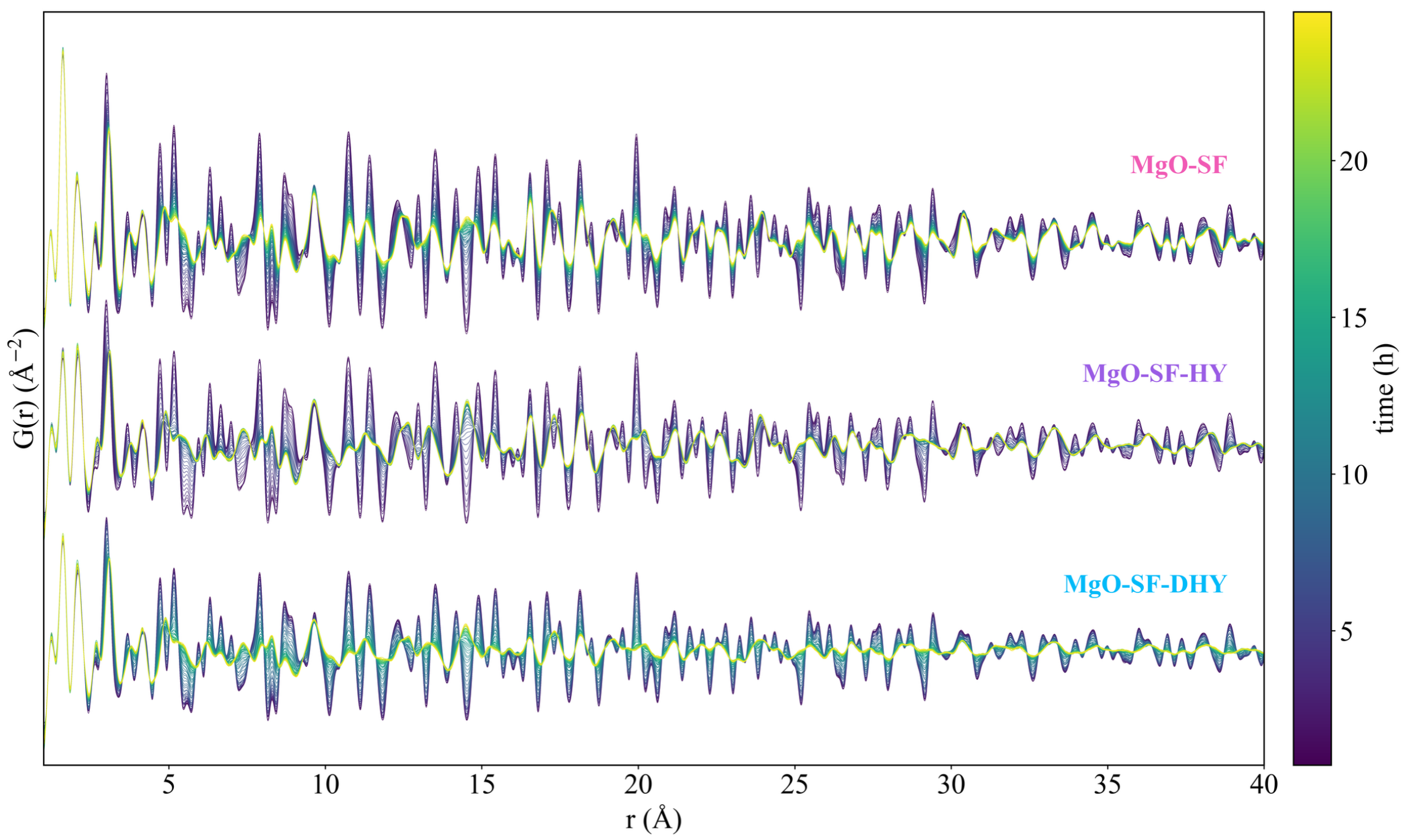

**Figure S2.** In situ X-ray PDFs of MgO-SF (top), MgO-SF-HY (middle), and MgO-SF-DHY (bottom) samples from 1-40 Å.

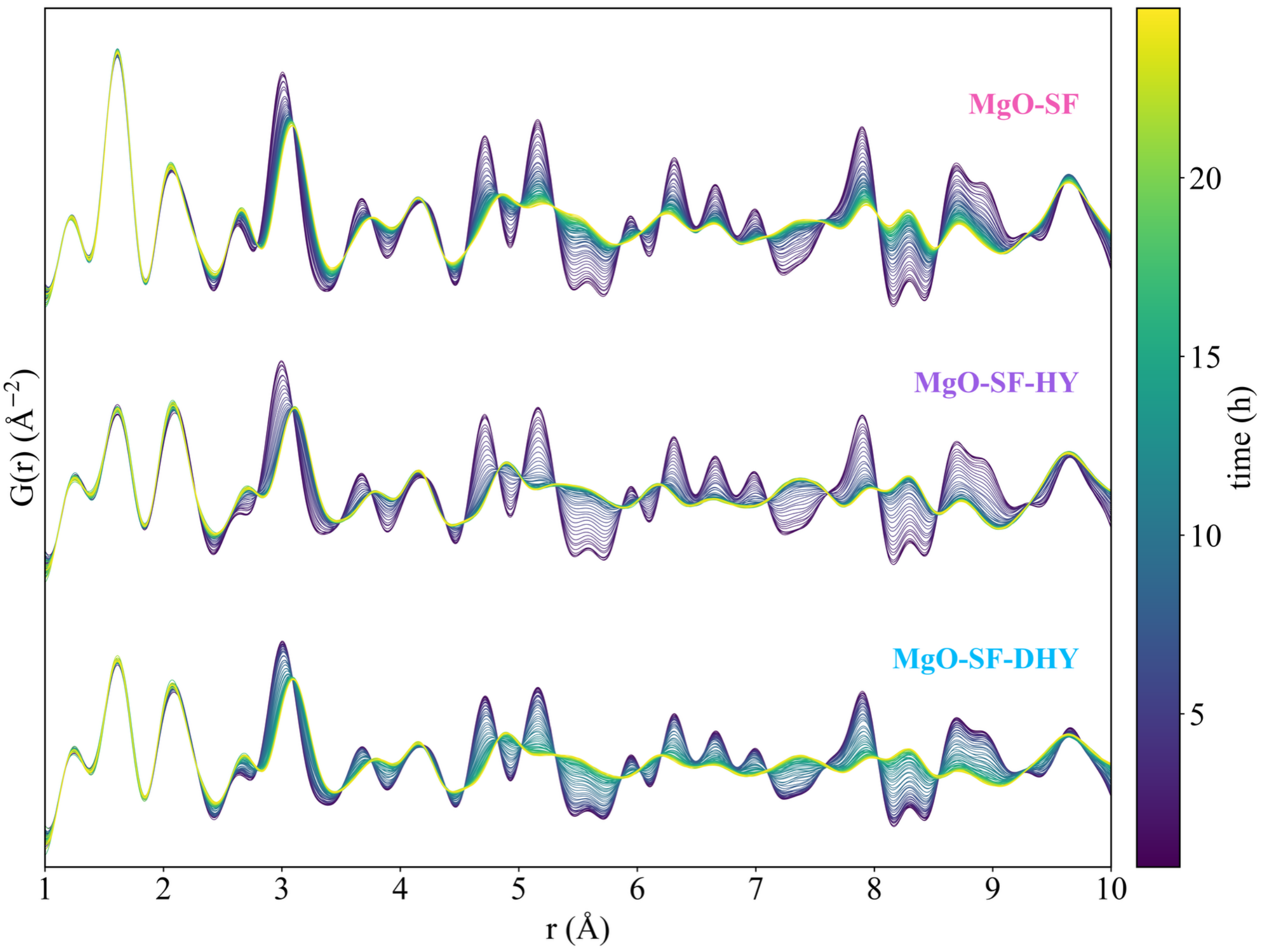

**Figure S3.** In situ X-ray PDFs of MgO-SF (top), MgO-SF-HY (middle), and MgO-SF-DHY (bottom) samples from 1-10 Å.

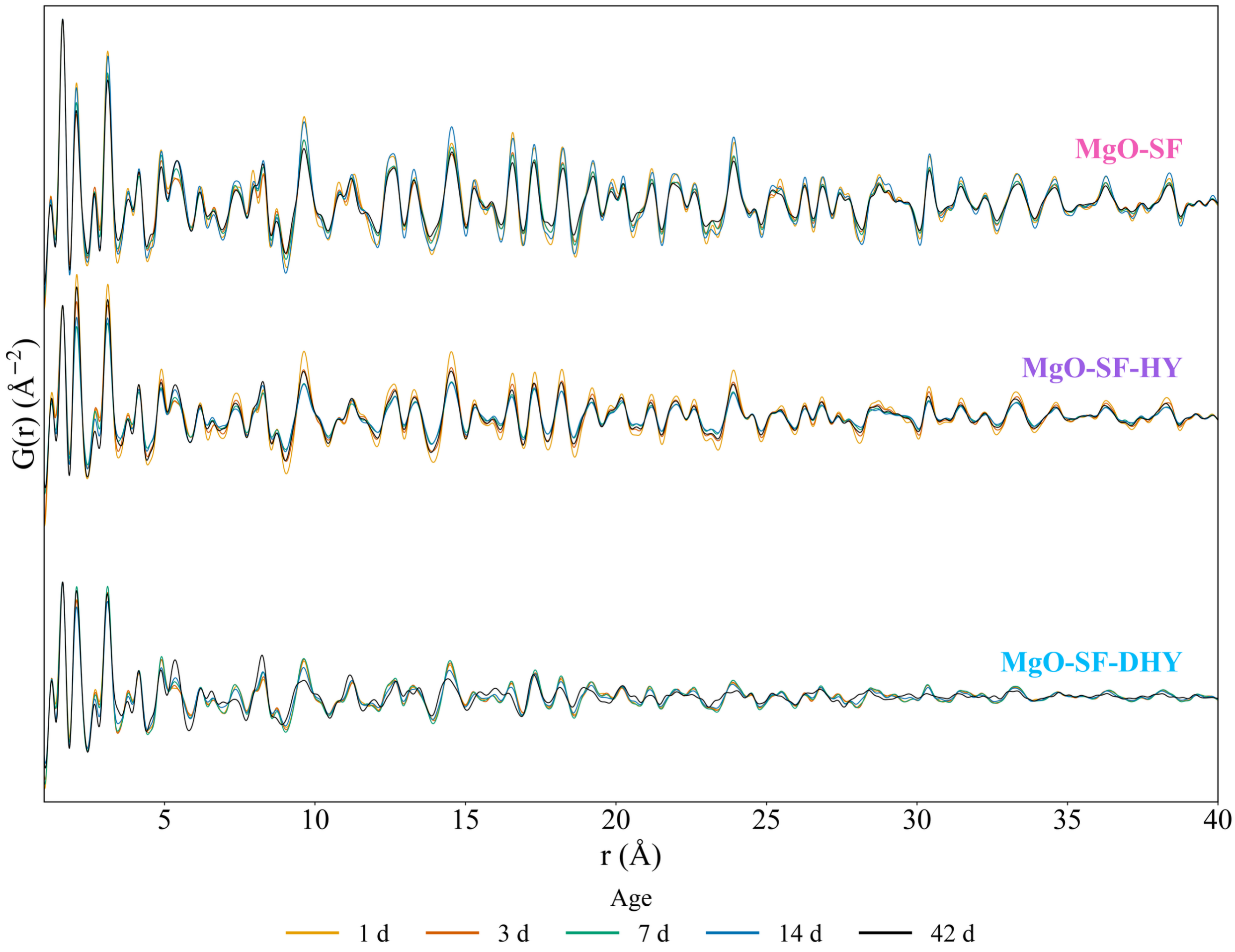

**Figure S4.** Ex situ X-ray PDFs of MgO-SF (top), MgO-SF-HY (middle), and MgO-SF-DHY (bottom) samples from 1-40 Å.

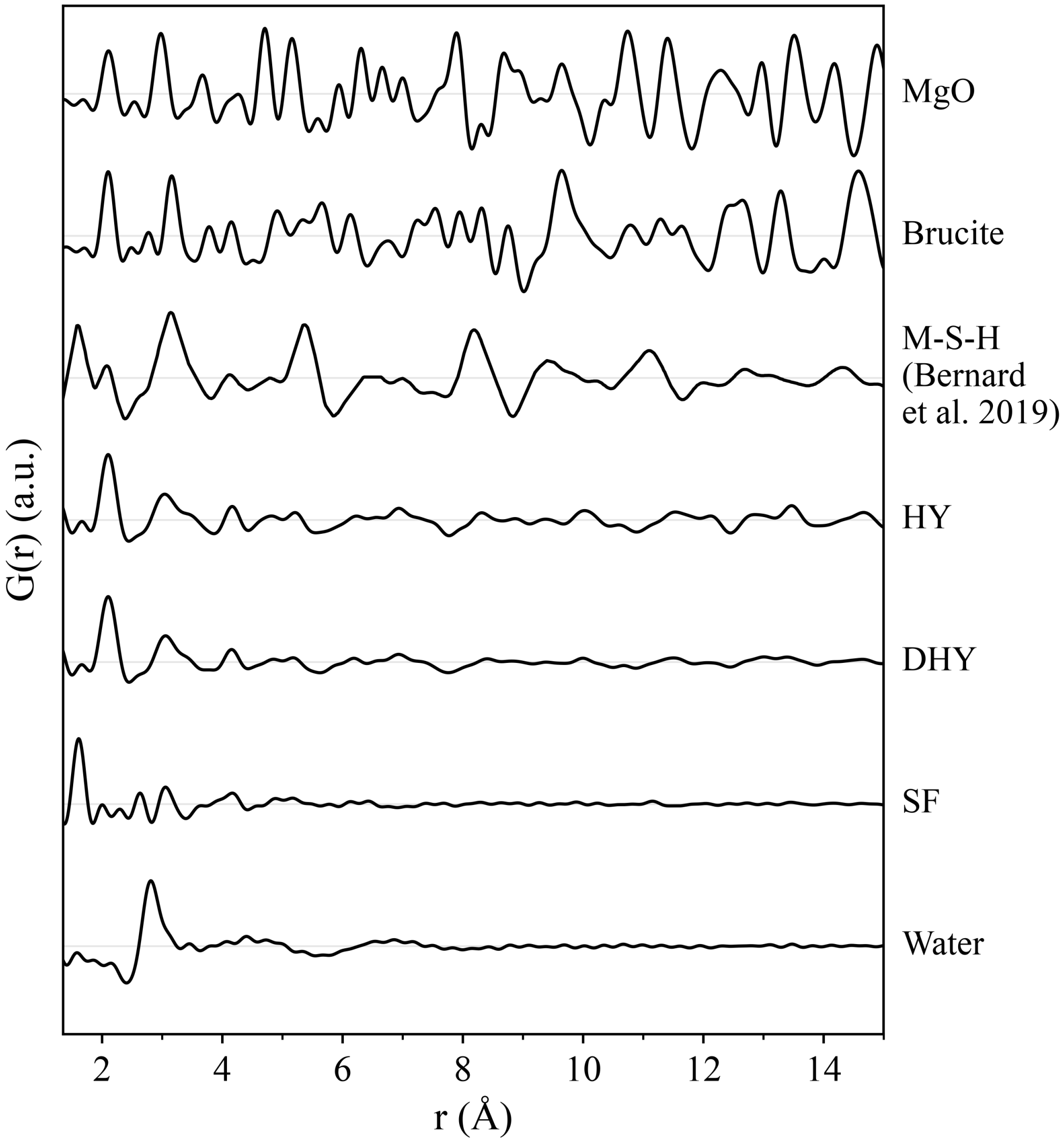

**Figure S5.** Reference PDFs of individual phases from 1.35-15 Å, used as fit components for real space phase quantification. PDFs of MgO and brucite were refined (see Materials and methods and Table S1 for more details on the refinement approach), M-S-H was extracted from Bernard et al.,[39] and HY, DHY, SF, and water were of experimental samples measured on the beamline. Each curve is normalized to its maximum peak height over the plotted range.

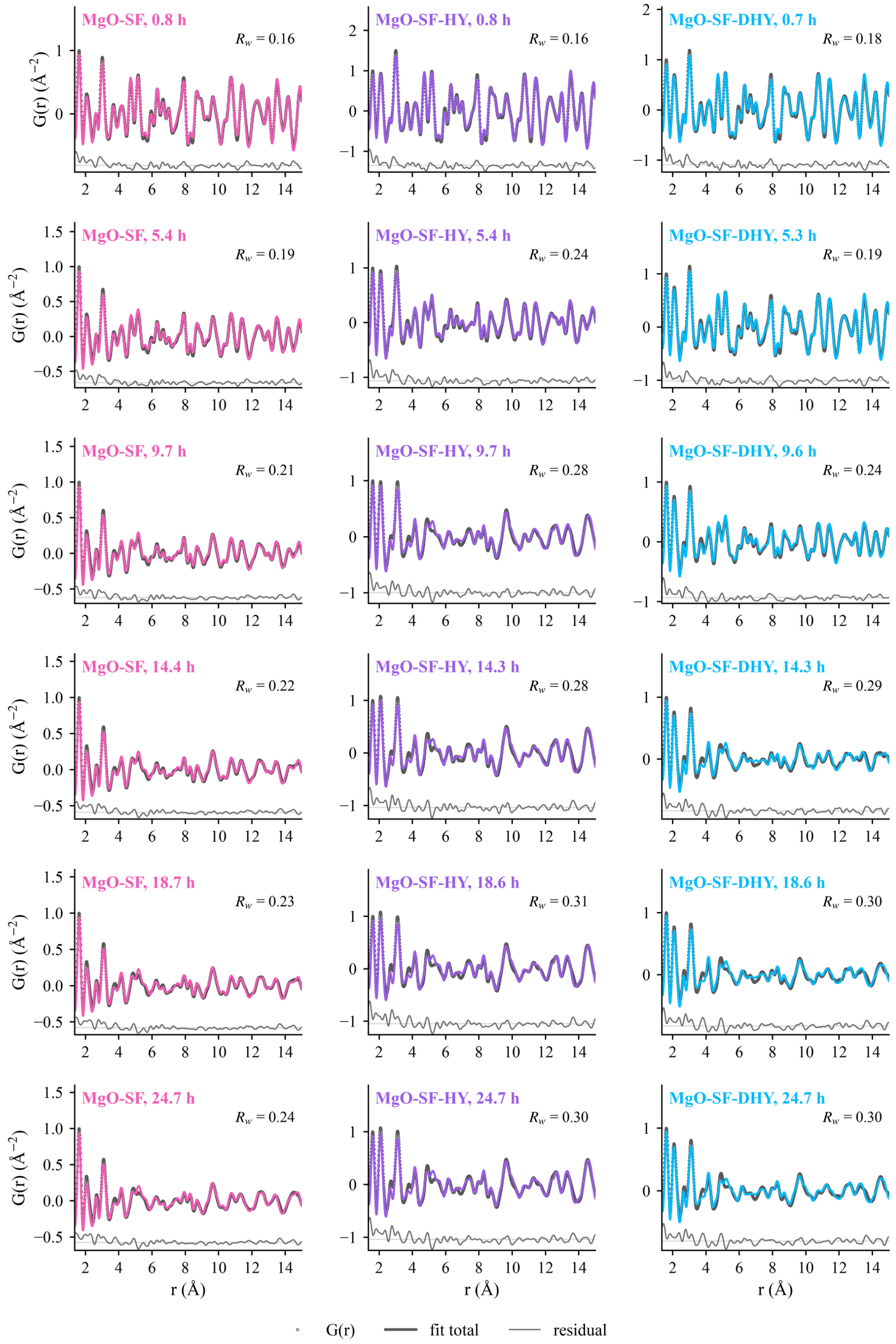

**Figure S6.** Real space phase quantification results for MgO-SF, MgO-SF-HY, and MgO-SF-DHY samples, showing select in situ PDF fits from 1.35-15 Å including the raw data (grey dots), fitted data (pink/purple/blue line), and residual (grey line). The goodness-of-fit ($R_w$) values are also reported.

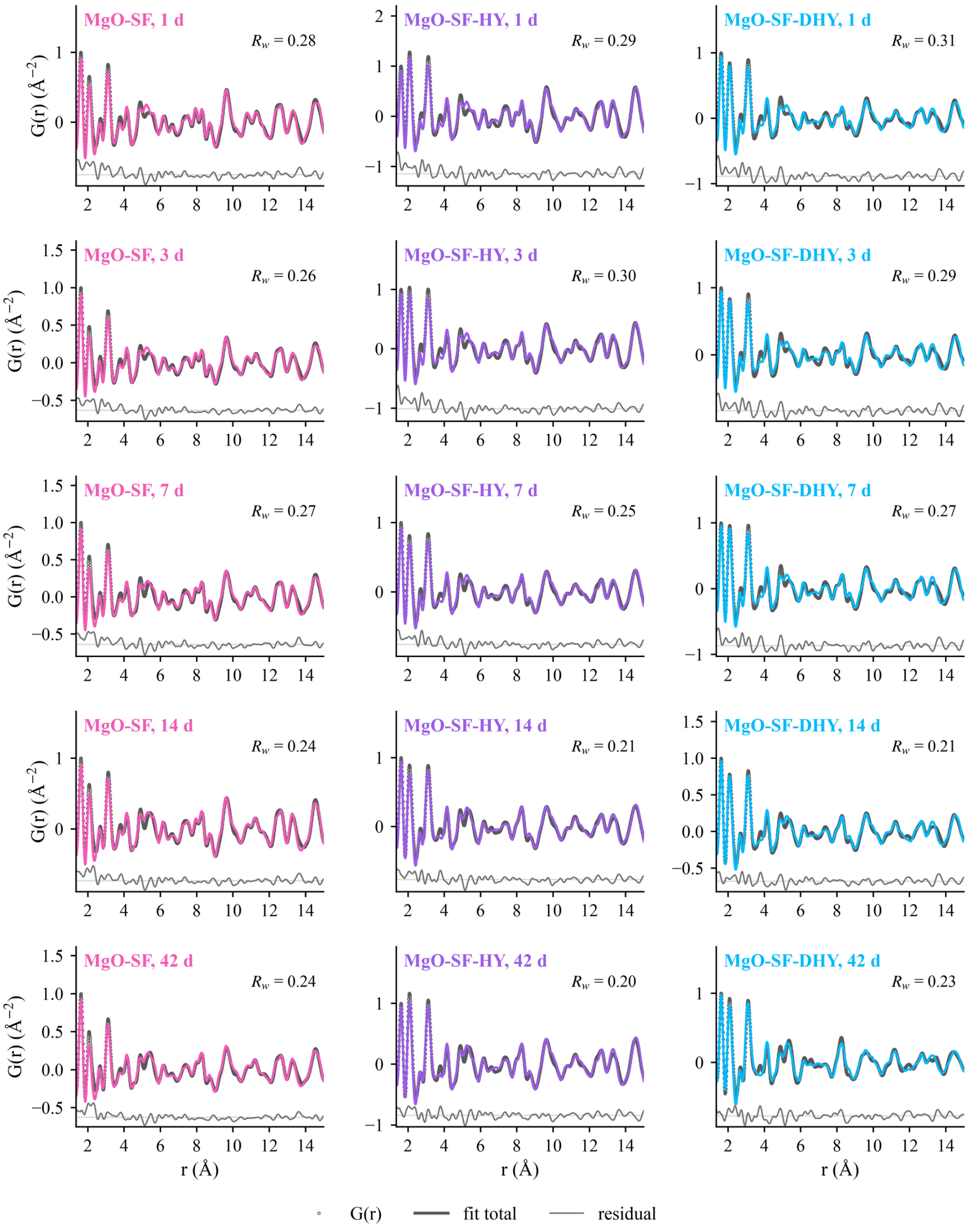

**Figure S7.** Real space phase quantification results for MgO-SF, MgO-SF-HY, and MgO-SF-DHY samples, showing all ex situ PDF fits from 1.35-15 Å including the raw data (grey dots), fitted data (pink/purple/blue line), and residual (grey line). The goodness-of-fit ($R_w$) values are also reported.

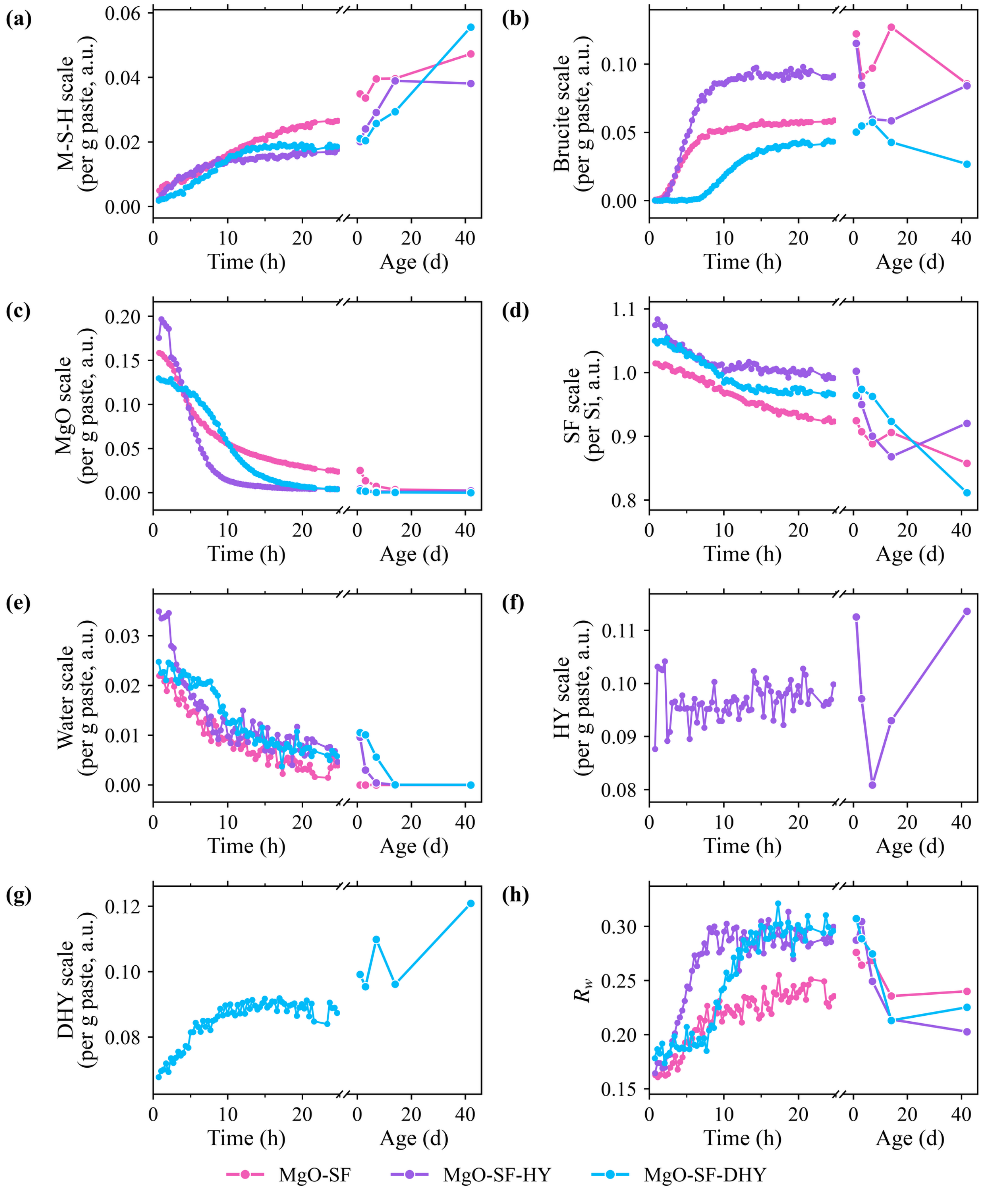

**Figure S8.** Scale factors for (a) M-S-H, (b) brucite, (c) MgO, (d) SF, (e) water, (f) HY, (g) DHY, and (h) $R_w$ values obtained from real space phase quantification, for in situ and ex situ PDFs based on fitting from 1.35-15 Å. Before fitting, each experimental PDF was normalized to the height of its own Si-O correlation (~1.6 Å), so the refined scale factors are proportional to the amount of each phase per Si in the paste. The scale factors in (a,b,c,e,f,g) are shown per g of paste; since all Si originates from SF, each per-Si scale factor was multiplied by the SF mass fraction of that paste's mix design (Table 1). Scale factor values are comparable between pastes for a given phase, however, these values are not comparable across different phases within a single paste. Note that for the MgO-SF-HY sample the scale of the HY phase was refined, while for the MgO-SF-DHY sample refinement of the DHY scale factor was carried out. Attempts to refine both HY and DHY for the MgO-SF-HY sample yielded a scale factor of ~0 for the DHY phase over the duration of the in situ and ex situ measurements. Similarly, the HY phase scale factor for the MgO-SF-DHY sample was found to be ~0. The increase in DHY scale factor in the MgO-SF-DHY sample during the initial ~10 h is attributed to rehydration of the DHY phase, as opposed to an increase in amount of DHY present, since the PDFs of DHY and HY (see Figure S5) are very similar, with the main difference being the higher intensity of the PDF peaks in HY compared with DHY.

**Table S1.** Refined parameters for MgO (refined once per sample using the first in situ scan) and brucite (refined once per sample using the final in situ scan at which point a sizable amount of brucite/HCB is present).

| **Phase** | **Sample** | **Scan time (h)** | **a (Å)** | **c (Å)** | **$U_{iso}$(Mg) ($Å^2$)** | **$U_{iso}$(O) ($Å^2$)** | **$\delta_2$ ($Å^2$)** |
|---|---|---|---|---|---|---|---|
| MgO | MgO-SF | 0.81 | 4.2157 | - | 0.0046 | 0.0054 | 2.35 |
| | MgO-SF-HY | 0.76 | 4.2166 | - | 0.0046 | 0.0053 | 1.75 |
| | MgO-SF-DHY | 0.71 | 4.2179 | - | 0.0046 | 0.0052 | 1.54 |
| Brucite | MgO-SF | 24.74 | 3.1405 | 4.7738 | 0.0057 | 0.0097 | 3.66 |
| | MgO-SF-HY | 24.70 | 3.1415 | 4.7642 | 0.0061 | 0.0109 | 3.01 |
| | MgO-SF-DHY | 24.65 | 3.1375 | 4.7641 | 0.0052 | 0.0112 | 3.65 |